\documentclass[ijms,review,accept,pdftex,oneauthor]{Definitions/mdpi} 
\usepackage[version=4]{mhchem}
\usepackage{tikz}
\usepackage{makecell}  
\usepackage{booktabs}  
\usepackage{tabularx}
\usepackage{adjustbox}
\usepackage{pifont}

\usepackage[
    starfontserif 
    ]{starfont}
\usepackage{amsmath}

\DeclareSymbolFont{starfontsym}{OT1}{sts}{m}{n}
\DeclareMathSymbol{\mathSun}{\mathord}{starfontsym}{115}
\DeclareMathSymbol{\mathMercury}{\mathord}{starfontsym}{102}
\DeclareMathSymbol{\mathVenus}{\mathord}{starfontsym}{103}
\DeclareMathSymbol{\mathTerra}{\mathord}{starfontsym}{76}
\DeclareMathSymbol{\mathvarTerra}{\mathord}{starfontsym}{108}
\DeclareMathSymbol{\mathMoon}{\mathord}{starfontsym}{100}
\DeclareMathSymbol{\mathvarMoon}{\mathord}{starfontsym}{97}
\DeclareMathSymbol{\mathMars}{\mathord}{starfontsym}{104}
\DeclareMathSymbol{\mathJupiter}{\mathord}{starfontsym}{106}
\DeclareMathSymbol{\mathSaturn}{\mathord}{starfontsym}{83}
\DeclareMathSymbol{\mathUranus}{\mathord}{starfontsym}{70}
\DeclareMathSymbol{\mathvarUranus}{\mathord}{starfontsym}{65}
\DeclareMathSymbol{\mathNeptune}{\mathord}{starfontsym}{71}
\DeclareMathSymbol{\mathPluto}{\mathord}{starfontsym}{74}
\DeclareMathSymbol{\mathvarPluto}{\mathord}{starfontsym}{72}

\firstpage{1} 
\pubvolume{26}
\issuenum{1}
\articlenumber{7531}
\pubyear{2025}
\copyrightyear{2025}
\externaleditor{   } 
\datereceived{12 June 2025} 
\daterevised{26 July 2025} 
\dateaccepted{1 August 2025} 
\datepublished{4 August 2025} 
\hreflink{https://doi.org/10.3390/ijms26157531} 

\Title{Photochemical Haze Formation on Titan and Uranus: A Comparative Review}

\TitleCitation{Photochemical Haze Formation on Titan and Uranus: A Comparative Review}

\Author{David Dubois 
 $^{1,2}$\orcidA{}}

\AuthorNames{Dubois, D.}

\isAPAStyle{%
       \AuthorCitation{Lastname, F., Lastname, F., \& Lastname, F.}
         }{%
        \isChicagoStyle{%
        \AuthorCitation{Lastname, Firstname, Firstname Lastname, and Firstname Lastname.}
        }{
        \AuthorCitation{Dubois, D.}
        }
}

\address{%
$^{1}$ \quad NASA Ames Research Center, MS 245-6, Moffett Field, CA 94035-1000
, USA; david.f.dubois@nasa.gov\\
$^{2}$ \quad Bay Area Environmental Research Institute, Moffett Field, CA 94035
, USA\\}

\abstract{The formation and evolution of haze layers in planetary atmospheres play a critical role in shaping their chemical composition, radiative balance, and optical properties. In the outer solar system, the atmospheres of Titan and the giant planets exhibit a wide range of compositional and seasonal variability, creating environments favorable for the production of complex organic molecules under low-temperature conditions. Among them, Uranus---the smallest of the ice giants---has, since Voyager 2, emerged as a compelling target for future exploration due to unanswered questions regarding the composition and structure of its atmosphere, as well as its ring system and diverse icy moon population (which includes four possible ocean worlds). Titan, as the only moon to harbor a dense atmosphere, presents some of the most complex and unique organics found in the solar system. Central to the production of these organics are chemical processes driven by low-energy photons and electrons (<50 eV), which initiate reaction pathways leading to the formation of organic species and gas phase precursors to high-molecular-weight compounds, including aerosols. These aerosols, in turn, remain susceptible to further processing by low-energy UV radiation as they are transported from the upper atmosphere to the lower stratosphere and troposphere where condensation occurs. In this review, I aim to summarize 
the current understanding of low-energy (<50 eV) photon- and electron-induced chemistry, drawing on decades of insights from studies of Titan, with the objective of evaluating the relevance and extent of these processes on Uranus in anticipation of future observational and \textit{in situ} exploration.}


\begin{document}




\section{Introduction}
{In the four decades since Voyager 1 and Voyager 2 swept past all four gas giants, modelers, astronomers, and~experimentalists have been incentivized to come one step closer to examining the outer planets and, ultimately, Uranus and Neptune.} 
While the Jovian and Kronian systems have been receiving increased scrutiny, Uranus and Neptune remain the final frontiers of planetary exploration in our solar system. Recognized as a flagship mission top priority for the next decade, the~selection of the Uranus Orbiter Probe (UOP) by the \textit{Decadal Strategy for Planetary Science and Astrobiology 2023--2032
} \citep{Decadal20232032} illustrates a renewed interest in the next \ce{H2}-dominated planet. This window offers an opportunity to utilize decades of knowledge obtained from Jupiter to Saturn and the moons of these planets, as~well as instrument development which could prove useful to characterize the atmospheres of Uranus or Neptune. As~defined in \citep{Decadal20232032}, UOP would target fundamental questions about the origin and evolution of our solar system and the dynamics of chemical disequilibria across the atmosphere (Q7.1c), the~chemical and physical processes influencing haze formation (Q7.3d), local ion/neutral composition (Q7.4d), and~seasonal effects triggering atmospheric chemistry and variability in haze production (Q7.5c). For~the latter topic, supportive laboratory and numerical studies of high- and low-pressure chemistry, reaction rates, and~photochemistry will especially be relevant to improve our understanding of Uranus' atmospheric dynamics and~chemistry.

Closer than Uranus, Saturn's largest moon Titan is the only moon in the solar system with a dense atmosphere. This atmosphere contains numerous organic molecules as well as detached haze layers resulting from intense photochemical activity in the upper atmosphere. The~opaque organic haze hides a geologically rich surface made of dunes, craters, and~liquid methane and ethane lakes. The~Cassini-Huygens mission which arrived at Titan in 2004 (and Huygens landing on Titan in 2005), unmasked for 13 years several of Titan's characteristics until the end-of-mission in 2017. After~almost 350 years since its discovery by Christiaan Huygens in 1655, many of Titan's secrets were about to be unveiled 
 \citep{Lebreton1992, Lebreton2002,Lebreton2005,Brown2009,Lopes2025TitanCassini-Huygens}. 
The unprecedented data collected by Cassini and Huygens at Titan unveiled, over~a large spectrum covering the UV up to radio wavelengths, many of the atmosphere's physical, chemical, thermal, and~transport characteristics from the surface to the thermosphere. The~Cassini-Huygens mission has deepened our understanding of the role of photochemistry on haze formation, benefiting from holistic experimental, modeling, and~observational studies \citep{Cable2012,Horst2017,Nixon2018,Nixon2024,Lopes2025TitanCassini-Huygens}. Investigations have shed light on the seasonal variations in temperature and cloud coverage, the~structure of the detached haze layer, and~the role of neutral-ion reactivity on the nucleation and growth of high-altitude aerosols. Still, many open questions prevail \citep{Nixon2018} and in addition to newer observations conducted to expand our knowledge of Titan's chemical inventory, Cassini's data heritage remains vast enough to probe for many years to come \citep{Nixon2016,Nixon2018,Coy2023}.

Across the four gas giant planets' \ce{H2}-dominated atmospheres, photochemical process are mainly driven by ultraviolet (UV) photons, magnetospheric energetic electrons and auroral deposition, galactic cosmic rays (GCR), and~lightning-induced chemistry. The~chemical inventory in these atmospheres directly depends on the source and intensity of photon deposition, and~the photon penetration depth (see sections hereafter
). In~this context, the~focus of this review will be on studies that have explored photochemical processes within three wavelength regions spanning UV photon energies from 3.1 to 
50 eV (Table \ref{Table 1 - energy range}); this is a~photon region at the source of fundamental dissociative and ionizing processes, resulting in the molecular and atmospheric diversity observed on Titan and Uranus. Region 1 contains Far-Ultraviolet (FUV) photons, Region 2 corresponds to Lyman-$\alpha$ radiation, and~Region 3 contains more energetic Extreme-Ultraviolet (EUV) photons. Studies investigating processes induced by low-energy (<50 eV) chemistry will be reviewed according to the wavelength range as classified in Table~\ref{Table 1 - energy range}. 

\begin{table}[H]
\caption{Spectral regions with associated energy (eV) and wavelength~ranges. \label{Table 1 - energy range}}
\begin{tabularx}{\textwidth}{lCCC}
\toprule
 & \textit{\textbf{Region 1
}} & \textit{\textbf{Region 2
}} & \textit{\textbf{Region 3
}}\\
 \midrule
 & \makecell{\textbf{Near/Far-UV
}\\(400--121.6 nm)}	
 & \makecell{\textbf{Lyman-$\alpha$
}\\(121.6 nm)}	
 & \makecell{\textbf{EUV/VUV
}\\(121.6--25 nm)} \\
\midrule
Energy range (eV) & 3.1--10.2 & 10.2 & 10.2--49.6 \\
\bottomrule
\end{tabularx}
\end{table}


Although the atmospheric compositions of both Titan and Uranus are distinct, decades of Titan research might help in identifying areas and techniques for both bodies needing further scrutiny into their atmospheric photochemical processes. Furthermore, the~second most abundant category of exoplanets in the most comprehensive catalogues consists of mini-Neptune and Neptune exoplanets. These exoplanets have sizes of $2<R_{\mathTerra}<6$, but~the definition of this size classification itself relies on planets (i.e.,~Uranus and Neptune) whose characteristics themselves are poorly known. As~such, further exploring Uranus (and Neptune) will not only help in exoplanetary characterization, but~also in our understanding of planetary formation and the solar system's evolution. Within~our solar system, all four~hydrogen-helium-rich planets hold most of the planetary mass of the solar system. Both ice giants hold a combined total of 41 moons. Our understanding of the formation and evolution of these systems relies on planetary formation and hydrodynamic \mbox{models \citep{Goldreich2004PlanetNeptune,Helled2014TheExoplanets}}.


This review is structured into four 
sections as follow:

\begin{itemize}
    \item Section 
 \ref{Section 2} will include a review of the atmospheric and radiative environments of the ice giants with an emphasis on Uranus and how they differ from Titan in the Saturnian system. In~addition, I will survey the current state-of-the-art body of knowledge of their chemical inventories.

    \item 
 Section \ref{Section 3} will include a review of the state of knowledge 
 of low-energy (<50 eV) photochemistry-induced mechanisms on Titan and Uranus, covering observations and \textit{in~situ} measurements, experimental simulations of gas phase and condensed state chemistry, photochemical modeling, dicationic chemistry, and~recent advances in quantum chemical calculations. Important aspects of branching ratio determination will also be discussed.
    \item 
 Section \ref{Section 4} will be dedicated to negative ions. 
 Discoveries pertaining to negative ion chemistry on Titan deserve their own section, as~their participation in molecular and haze growth requiring photochemical and radiative processes has proven to \mbox{be substantial.}
    \item 
 Section \ref{Section 5 conclusions} will conclude with a summary of potential future investigations needed to probe Uranus and prepare for upcoming studies before future missions to the gas giants.

\end{itemize}

\section{The Chemical and Radiative Environments of Titan and the Ice~Giants}
\label{Section 2}
\unskip
\subsection{Ice Giants: General~Considerations}
\label{Section 2.1}

\begin{figure}[H]

\begin{adjustwidth}{-\extralength}{0cm}
\centering 
\includegraphics[width=15 cm]{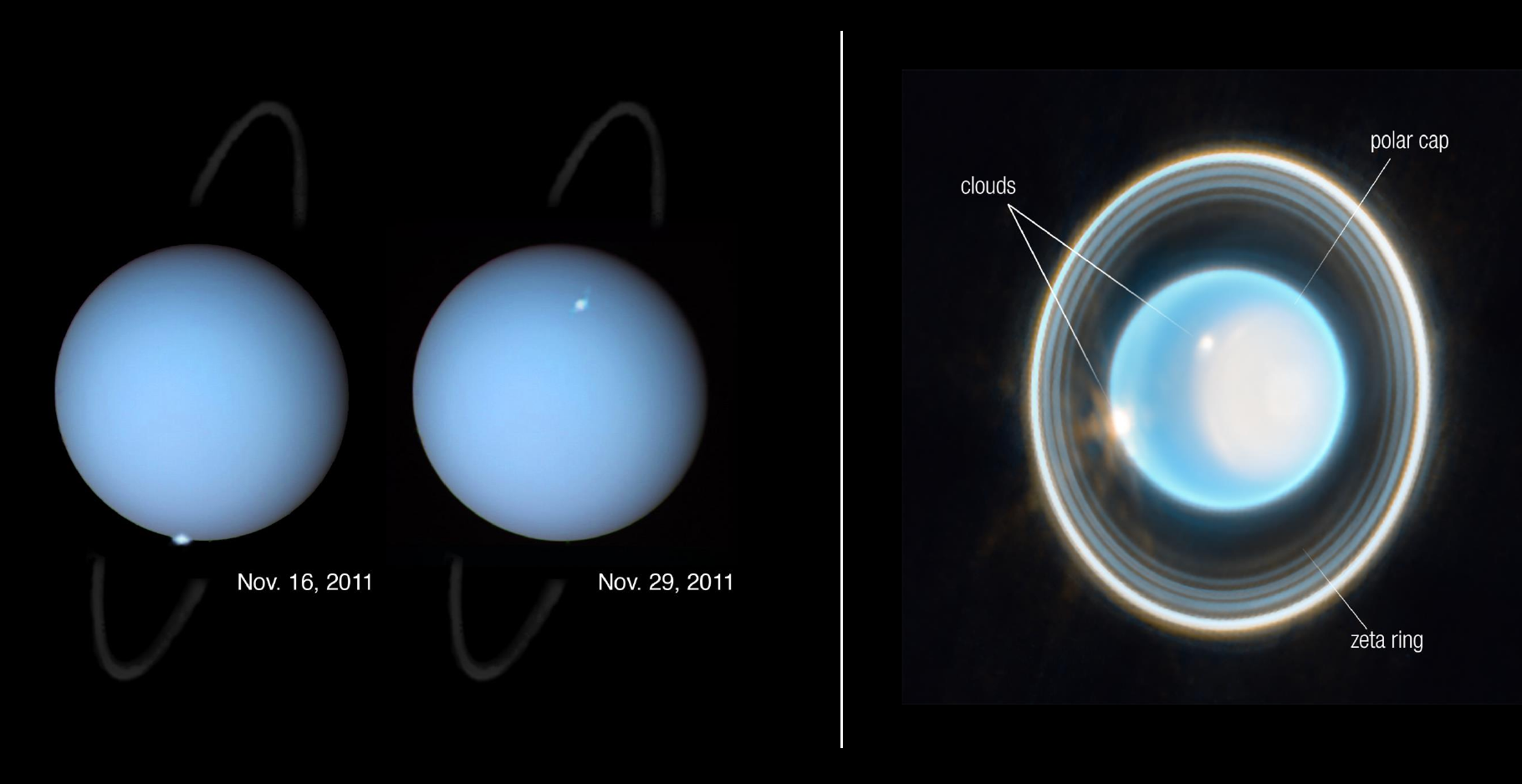}
\end{adjustwidth}
\caption{\textbf{Left
}: Composite image of Uranus combining auroral observations as seen in FUV by HTS/STIS for the first time in November 2011 \citep{Lamy2012Earth-basedAurorae}, with~the planet's blue disk seen by Voyager 2 in 1986. The~faint ring system observed by the Gemini Observatory observed in 2011 is also overlaid. \textbf{Right}: JWST/NIRCam image taken on February 6, 2023. Bright sun-facing polar cap and multiple bright cloud systems can be seen. The~planet's complex 13-ring system was also captured by NIRCam with the innermost and optically thin $\zeta$ ring being visible. Image credits: NASA, ESA, and~L. Lamy (Observatory of Paris, CNRS, CNES); NASA, ESA, CSA, STScI, Joseph DePasquale (STScI).\label{Figure 1 Uranus Voyager}}
\end{figure}

Uranus and Neptune (the ``ice giants'') are the least explored planets of our solar system. To~date, Voyager 2 is the only mission to have flown by anywhere close to Uranus and Neptune, at~a distance of 4 $R_{\mathUranus}$ and $1/5$ $R_{\mathNeptune}$, respectively. For~the first time, these missions revealed unprecedented imagery and data of not just the two farthest gaseous planets, but~also 16 new moons and 2 new rings (Figure \ref{Figure 1 Uranus Voyager}). Since Voyager 2 departed Neptune's vicinity in 1989, no other dedicated spacecraft has gone back.
The ice giants stand apart from Jupiter and Saturn in their own category. On~a compositional level, Uranus and Neptune both contain multiple suspected cloud layers stratified in \ce{CH4}, \ce{H2S}, \ce{NH4SH}, \ce{H2O}, and~\ce{NH3}. The~temperature profiles in these planets are poorly constrained and usually assumed to follow a standard moist adiabatic and well-defined profile \citep{Guillot2019UranusAtmospheres}. Radio occultation and mid-infrared measurements supported by global-average thermochemical equilibrium modeling helped in characterizing the atmospheric regions of these \mbox{planets \citep{Moses1992NucleationAtmosphere,Orton2014Mid-InfraredStratosphere,Moses2018SeasonalNeptune,Hueso2019AtmosphericLayers,Moses2020AtmosphericNeptune,Fletcher2021TheUranus}}. 

\begin{figure}[H]

\begin{adjustwidth}{-\extralength}{0cm}
\centering 
\includegraphics[width=15.5 cm]{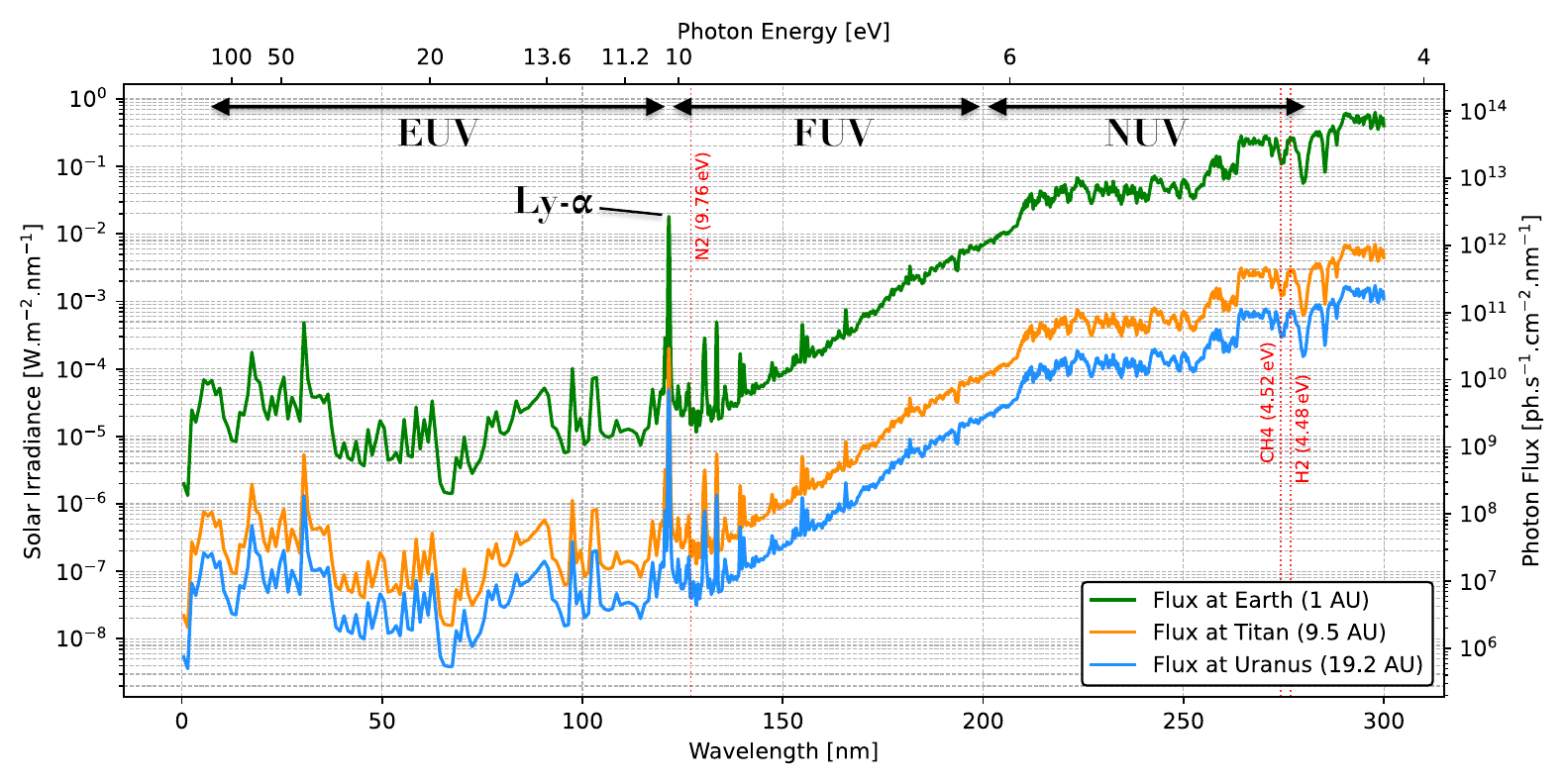}
\end{adjustwidth}
\caption{Disk-
 integrated high-resolution solar spectrum irradiance measured by SOLAR-HRS at Earth (1 AU) in green shown between 0.5 nm and 300 nm (data obtained with permission from \citep{Meftah2023TheCases}). The~data, measured in 2022, represents a reference of a solar minimum spectrum with a spectral resolution \mbox{<1.0 nm}. The~spectrum is also scaled at Titan (9.5 AU) in orange and Uranus (19.2 AU) in blue here for comparison, decreasing at a scale of $1/r^2$. The~intense Ly-$\alpha$ band can be seen at 121.6 nm, and~red dotted lines correspond to the dissociative energy thresholds for \ce{N2}, \ce{CH4}, and~\ce{H2}. \label{Figure XX SOLAR-SSI-HRS}}
\end{figure}

A deep troposphere resides below the 0.1 bar pressure level, where temperatures are predicted to range from $\sim$200 K at the 10 bar level down to 50--60 K at 0.1 bar. Marking the tropopause at 0.1 bar (50 km) in both planets, the~temperature inversion sets the boundary between the troposphere and the stratosphere wherein temperatures substantially increase with increasing altitude up to the thermospheric boundary located at the 1 $\upmu$bar level (\mbox{550 km}). At~these altitudes, Uranus is already much warmer than Neptune (\mbox{350 K} vs. \mbox{180 K}). In~the upper atmosphere extending up to 6000 km in the case of Uranus, Voyager 2 measurements taken by the Ultraviolet Spectrometer (UVS) show temperatures reaching 750 K near the exobase \citep{Melin2020TheNeptune}. The~Neptunian exobase is located around \mbox{4000 km} \citep{Broadfoot1989UltravioletTriton}. Notwithstanding large distances from the Sun, both upper atmospheres are hotter than Saturn's upper atmosphere, while Neptune's is almost as hot as Jupiter's, when considering solar heating alone \citep{Melin2020TheNeptune}. Interestingly, Uranus and Neptune have both undergone an important cooling phase of their upper atmospheres since Voyager's flybys \citep{Melin2020TheNeptune,Moore2020AtmosphericNeptune}. Energetics and dynamics mechanisms (high-energy particle ionization, auroral processes, photoionization, solar and seasonal cycles) have been investigated to understand the intense cooling of Uranus's atmosphere but have fallen short of combining all the complex seasonal, photochemical, orbital, and~dynamical processes shaping into one cohesive theory (\citep{Melin2020TheNeptune}, \textit{and references therein
}). While both planets share similar temperature and atmospheric profiles, they also exhibit distinct additional characteristics. Vertical mixing is, in~the standard models, assumed to be negligible at Uranus, while much more vigorous mixing is present at Neptune \citep{Strobel1991TheUranus,Guillot2019UranusAtmospheres,Moses2020AtmosphericNeptune}. Eddy diffusion (\textbf{$K_z$}) has been described as ``sluggish'' by \citep{Atreya1991PhotochemistryMixing} with equatorial values at the homopause ranging from 3000 to 10,000 cm$^2$ s$^{-1}$ \citep{Orton1990CalibrationNeptune,Atreya1991PhotochemistryMixing,Encrenaz1998ISOCoefficient,Orton2014Mid-InfraredStratosphere,Moses2018SeasonalNeptune}. Earlier modeling had placed vertical mixing even lower <100 mbar at \mbox{200 cm$^2$ s$^{-1}$}. Eddy diffusion is the main transport mechanism for vertical mixing but its coefficient is an important free parameter in 1D photochemical models. However, the~assumption of a temperature profile which follows a moist adiabatic lapse rate does not fully reproduce the likelihood that latitudinal variations due to meridional circulation, super-adiabatic conditions, and~turbulent updrafts/downdrafts may strongly influence the distribution of photochemically produced species in the atmosphere \citep{Guillot2019UranusAtmospheres,Fletcher2021TheUranus}. Coupled to auroral processes and auroral driven Joule heating, and~EUV photoionization, the~unique orbital geometry of Uranus (i.e.,~an extreme axial tilt of 97.8$^{\circ}$) makes it a unique planet in the solar system. Although~far from the Sun, solar spectrum irradiance (SSI) may reach $\sim$10$^{-5}$--10$^{-4}$ W m$^{-2}$ nm$^{-1}$ at Lyman-alpha (Ly-$\alpha$) wavelengths at the top of Uranus's atmosphere (Figure \ref{Figure XX SOLAR-SSI-HRS}). In addition, Uranus does not have an internal heating source, unlike Neptune 
 \citep{Pearl1990TheData}, and~the lack of understanding in the radiant energy budget of Uranus has made elucidating the planet's weather and dynamical patterns difficult. However, recent modeling studies spanning the 
 entire orbital period of Uranus have shed light on its internal heat flux and shown that Uranus does, in~fact, 
 possess a relatively significant internal heat source. This internal heat-to-absorbed 
 solar power ratio is 
 still much lower than that of the other gas giants \citep{Wang2025InternalUranus}. More future studies are 
 needed to resolve these discrepancies between models and observations \citep{Melin2020TheNeptune}.

\subsection{The Atmosphere of~Uranus}

The atmosphere of Uranus has a mean molecular weight of around 2.3 which increases to 3.1 deeper in the troposphere below 1 bar where \ce{CH4} reaches a near-constant mixing ratio of $\sim$3\% \citep{Guillot2019UranusAtmospheres,Hueso2019AtmosphericLayers,Sromovsky2019The2015}. The~intensity of solar EUV across the solar system decreases as the square of solar distance, where we find a solar constant of $\sim$14.8 W m$^{-2}$ at the Saturnian system and 3.7 W m$^{-2}$ at Uranus (Table \ref{Table2 planet parameters}). EUV intensity in the 90--110 nm range (11.27--13.78 eV, Region 3, Table~\ref{Table 1 - energy range}) falls at 0.21 kR and 0.05 kR, respectively. Presented differently, Voyager 2 UVS measurements of FUV reflectance at the Uranian subsolar point (i.e.,~close to the rotation axis) by \citep{Yelle1987AnalysisUranus} helped constrain the \ce{CH4} column abundance even before the closest approach. At~the same time, occultation experiments helped quantify the column abundances of \ce{H2}, \ce{H}, and~\ce{C2H2} \citep{Herbert1987The2}. These occultation measurements found that the opacity could be explained by the dominating presence of Rayleigh scattering of 
 \ce{H2} along with the presence of a Raman line at 1280 
 \text{\AA} and 
 \ce{C2H2} absorption near 1300 
 \text{\AA} and 1500 
 \text{\AA}  \citep{Yelle1987AnalysisUranus,Atreya1991PhotochemistryMixing}. More recently, Ref. \citep{Lamy2012Earth-basedAurorae} conducted the first Earth-based detection of an auroral event (transient emissions on the order of 1--2 kR) at Uranus, observed in the FUV in November 2011 during an intense solar wind event. These observations clearly indicated that there exist variations in the configuration between the solar wind and the magnetosphere, and~therefore the thermal and compositional structure of the upper atmosphere. Later measurements by the Hubble Space Telescope (HST) permitted a re-analysis of Voyager 2's albedo measurements from \citep{Yelle1989TheEncounter}, which brought down its value from $\sim$20--30\% to 5--10\% between 135 and 155 nm (8.0--9.19 eV, Region 1, Table~\ref{Table 1 - energy range}) \citep{Barthelemy2014DayglowBands}. The~emission features studied therein would have required an estimated energy flux of the precipitating electrons of 0.04--0.07 erg cm$^{-2}$ s$^{-1}$, but~sensitivity shortcomings prevented any determination of their actual energy in the 20 eV--20 keV range. Nonetheless, this first Earth-based detection of a transient auroral \ce{H2} emission came in good agreement with the Voyager 2-era disk average flux upper limit of 0.008 erg cm$^{-2}$ s$^{-1}$ by \citep{Waite1988SuperthermalElectroglow}. The~modeled \ce{H2} spectrum resulting from solar photons and precipitating electrons considered low-energy (centered at 20 eV) and higher energy (3 keV) electrons. The~calculated $\chi^2$ dependence of the precipitating flux showed little difference between these two energies and that one single transient event would have caused a total electron flux enhancement by a factor of 4--9 \citep{Barthelemy2014DayglowBands}. Overall, both Voyager 2 and Earth-based observations have contributed to a better understanding of auroral and photochemical events in the upper atmosphere of Uranus. However, we are still far from an accurate global picture involving complex latitudinal, temperature profile variations, hydrocarbon contributions, magnetospheric configuration, and~accurate knowledge of solar FUV/EUV fluxes \citep{Barthelemy2014SensitivityFlux}. As~such, a~more accurate description of these fluxes (extrapolating also into more energetic X-ray fluxes) will ultimately help in better characterizing exoplanet environments and their atmospheric composition \citep{France2013TheStars,Linsky2013ComputingStars,Barthelemy2014SensitivityFlux,Linsky2024InferringAtmospheres}.

\vspace{-6pt}
\unskip

\begin{table}[H]
\caption{Planetary parameters and solar activity measured by Voyager 2 at Saturn, Titan, and~Uranus.\label{Table2 planet parameters}}
	\begin{adjustwidth}{-\extralength}{0cm}
		\begin{tabularx}{\fulllength}{cCCCCC}
			\toprule
			\textbf{}	& \textbf{Mass ($M_{\mathTerra}$)}	& \textbf{Solar \mbox{Constant (W~m\textsuperscript{$-$2})}}      & \textbf{\mbox{EUV Intensity (kR)}, 90--110 nm} & \textbf{T (1 bar Level) \mbox{in Kelvin}} & \textbf{Mean \mbox{Molecular Weight}}\\
			\midrule
\multirow[m]{1}{*}{Titan}	& 0.023			& 14.8			 & 0.21 & 94 & 27.8\\
                   \midrule
\multirow[m]{1}{*}{Saturn}    & 95.2			& 14.8			 & 0.21 & 134--145 & 2.0\\
                   \midrule
\multirow[m]{1}{*}{Uranus}    & 14.5			& 3.7		 & 0.05 & 76--86 & 2.3\\
			\bottomrule
		\end{tabularx}
	\end{adjustwidth}
	\noindent{\footnotesize{See~\cite{Strobel1991TheUranus,Hueso2019AtmosphericLayers,Moses2020AtmosphericNeptune} for references.}}
\end{table}

The first recorded molecular detection in Uranus occurred in 1869 by \citep{Secchi1869ResultatsSolaires.}: \textit{``Dans le vert et dans le bleu il y a deux raies très-larges et très-noires
''}. Unbeknownst to the astronomer and priest that this detection was in fact that of molecular hydrogen, this was followed by observations of unresolved faint bands by \citep{Huggins1871IV.1871}. Then, \textit{``The remarkable absorption taking place at Uranus shows itself in six strong lines''
} \citep{Huggins1871IV.1871,Chinnici2019ContributionsSciences}. This first set of two observations acquired very faint bands, but~it was not until 1934 that the first observations firmly detected \ce{CH4} as being the first confirmed compound in the atmosphere \citep{Adel1934ThePlanets}. Eighteen years later, the~\ce{H2} ro-vibrational band at 8270 \text{\AA} was discovered \citep{Herzberg1952SpectroscopicNeptune}, and~nearly thirty-four years later, {accurate measurements of the \ce{H2}:\ce{He} mole fraction were conducted and trace species were discovered} 
by the International Ultraviolet Explorer \citep{Encrenaz1986AIUE} and Voyager 2 \citep{Tyler1986VoyagerSatellites}. At~present, 9 neutral molecules have been directly detected in the atmosphere of Uranus (see Table~\ref{Table 3 elemental comp} for an elemental composition comparison between Uranus and Titan).

\begin{table}[H]
\caption{Summary of all the directly detected neutral molecules in the atmospheres of Titan and Uranus \citep{Adel1934ThePlanets,Herzberg1952SpectroscopicNeptune,Encrenaz1986AIUE,Tyler1986VoyagerSatellites,Conrath1987TheMeasurements,Encrenaz2004FirstUranus,Burgdorf2006DetectionSpectroscopy,Moreno2017DetectionObservations,IrwinDetectionAtmosphere,Nixon2024}.  \label{Table 3 elemental comp}}
\small
\begin{adjustwidth}{-\extralength}{0cm}
\begin{tabularx}{\fulllength}{cm{2cm}<{\centering}cCccccc}
\toprule
\textbf{Atoms} & \multicolumn{2}{c}{\textbf{C}} & \multicolumn{2}{c}{\textbf{N}} & \multicolumn{2}{c}{\textbf{O}} & \multicolumn{2}{c}{\textbf{Other}} \\
\midrule
 & \textbf{Titan
} & \textbf{Uranus} & \textbf{Titan} & \textbf{Uranus} & \textbf{Titan} & \textbf{Uranus} & \textbf{Titan} & \textbf{Uranus} \\
\midrule
1 & \ce{CH4} & \ce{CH4} & \ce{HCN}, \ce{HNC}, \ce{CH3CN}, \linebreak \ce{HC3N}, \ce{C3H3N}, \linebreak \ce{C3H5N}, \ce{C4H3N} & - & \ce{H2O}, \ce{CO} & \ce{CO} & - & \ce{H2}, \ce{H2S} \\
\midrule
2 & \ce{C2H2}, \ce{C2H4}, \linebreak \ce{C2H6} & \ce{C2H2}, \ce{C2H6} & \ce{N2}, \ce{C2N2} & - & \ce{CO2} & \ce{CO2} & - & - \\
\midrule
3 & \ce{C3H2}, \ce{C3H4}, \linebreak \ce{C3H6}, \ce{C3H8} & \ce{C3H4} & - & - & - & - & - & - \\
\midrule
4 & \ce{C4H2} & \ce{C4H2} & - & - & - & - & - & - \\
\midrule
5 & - & - & - & - & - & - & - & - \\
\midrule
6 & \ce{C6H6} & - & - & - & - & - & - & - \\
\bottomrule
\end{tabularx}
\end{adjustwidth}
\end{table}

\begin{figure}[H]
\includegraphics[width=14 cm]{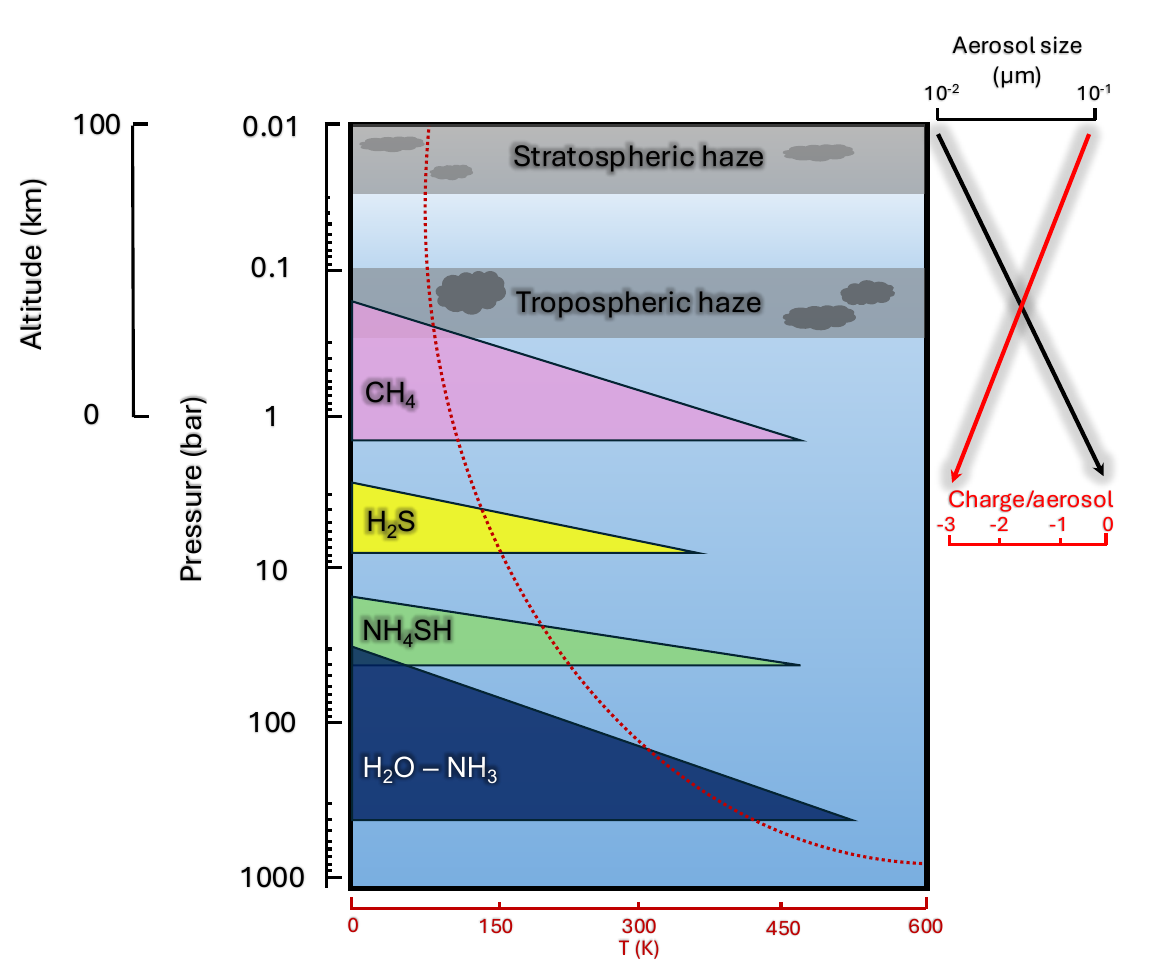}
\caption{Standard 
 picture schematic of the atmospheric profile of Uranus with the tropospheric and stratospheric haze layers, the~\ce{CH4}, \ce{H2S}, \ce{NH4SH}, and~\ce{H2O}$-$\ce{NH3} cloud decks, and~the approximate temperature profile \citep{Fegley1991SpectroscopyUranus,Guillot2019UranusAtmospheres,Hueso2019AtmosphericLayers,Moses2018SeasonalNeptune,Moses2020AtmosphericNeptune,Vorburger2024MassProbe}. The~general, approximate trends across the haze layers for aerosol particle size distribution (in $\upmu$m), and~the distribution of the mean charge per aerosol from \citep{Molina-Cuberos2023ThePlanets} at altitudes of 0.01--3 bar are also~given. \label{Figure 2 atm profile Uranus}}
\end{figure}

The atmosphere of Uranus starts with a deep troposphere which rises up to 1 bar, and~whose composition and structure are still, to~a large extent, poorly known \citep{Guillot2019UranusAtmospheres,Fletcher2021TheUranus}. \ce{CH4} is in abundance and drives the moist convection in the troposphere, before~it condenses due to the cold temperatures near the tropopause (1--1.5 bar). In~models assuming a temperature profile following a moist adiabat, all expected \ce{CH4}, \ce{H2S}, \ce{NH4SH}, and~\ce{H2O-NH3} species form stratified, well-defined cloud decks below 1 bar (Figure \ref{Figure 2 atm profile Uranus}). Methane, like all other species, display latitudinal and temporal variations and as a result are important indications of underlying seasonal and/or vertical transport \citep{Karkoschka2009TheSpectroscopy,Moses2018SeasonalNeptune,Fletcher2021TheUranus}. Indeed, as~explained in \citep{Fletcher2021TheUranus}, tropospheric updraft of methane and other hydrocarbons is unlikely given the cold trap at $\sim$50 K. However, mechanisms such as mid-latitude tropospheric updraft could in theory transport \ce{CH4}-rich air mass up into the stratosphere. Or, adiabatic warming may occur, thus moving subsiding air from the stratosphere into the troposphere \citep{Lunine1993TheNeptune,Moses2018SeasonalNeptune} (a similar seasonally driven stratospheric circulation was seen on Titan, discussed later). Modeling of the visible/near-infrared reflectivity of the haze profile on Uranus includes several uncertainties but was recently studied by \citep{Irwin2022HazySpots}. Their conclusion was that the atmosphere consists of three ``detached'' haze layers. (1) A vertically extended photochemical haze layer $<10^{-2}$ bar which is produced in the lower to mid stratosphere. Haze particle are then relatively mixed to lower altitudes by Eddy diffusion. (2) Right above the \ce{CH4} condensation altitude at $\sim$1 bar, haze particles of moderate size ($\sim$1~$\upmu$m) constitute a thin but highly opaque cloud structure composed of an important mixture of \ce{CH4} ice and photochemically produced particles, absorbent of UV and long-wavelength photons. It is at this level that ice/haze particles are expected to act as cloud condensation nuclei (CCN). (3) Finally, a~deeper and darker aerosol layer for $P$ > 5--7 bar would be consistent with the presence of an \ce{H2S}-based layer mixed with photochemically generated aerosols and other ices. The~modeled haze particles are predicted to scatter light at 500 nm while absorbing photons at longer wavelengths. Future experimental and numerical studies of the optical and physico-chemical properties of these particles would help in constraining the vertical structure of Uranus' atmosphere which remains largely uncertain. Support from telescopic observations is ongoing \citep{Sanchez-Lavega2023DynamicsObservations} and recent surveys using the Lowell Observatory and HST have highlighted the dependency of cloud brightness variations with \ce{CH4} distributions and orbital position \citep{Irwin2024ModellingNeptune}.

A number of species such as \ce{HCN}, \ce{PH3}, \ce{GeH4}, \ce{HCl}, and~\ce{CH3SH} expected to be in disequilibrium have been speculated from thermochemical modeling to survive in the upper troposphere, thus importantly bridging the deep troposphere composition with the stratosphere \citep{Moses2020AtmosphericNeptune}. There are however many uncertainties surrounding this chemistry such as: knowledge of the deep tropospheric temperature profile, ice nucleation at low temperature, kilobar-level and high-temperature reactions rates, diffusion effects, and~non-ideal gas behavior, all of which are factors of uncertainty that have been highlighted \mbox{in \citep{Moses2020AtmosphericNeptune}}.
Above the 1 bar pressure (Figure \ref{Figure 2 atm profile Uranus}), the~stratosphere extends up to $\sim$1 $\upmu$bar where temperatures rapidly reach 300 K, following the moist adiabat, and~finally reach 750 K at the exobase \citep{Moses2018SeasonalNeptune}. This vertical structure interpretation is almost certainly erroneous for the reasons mentioned previously, and~the reality is likely to involve many altitudinal, latitudinal, and~seasonal variations \citep{Guillot2019UranusAtmospheres}. These atmospheric parameters will regulate the distribution of species in the atmosphere, and~these species will also influence the cloud structure and haze formation of the planet. Therefore, it is crucial to obtain a better understanding of the chemical inventory and physico-chemical processes in~Uranus.

\vspace{-9pt}
\unskip

\subsection{Chemical Inventory in~Uranus}

This section will focus on the available inventory of detected species in the atmospheric column of Uranus. The~origin and photolytical evolution of the species are still largely unknown, and~as pointed out in \citep{Mousis2018ScientificExplorations} the differences observed with the other gas giants may result from different production/loss photolytic rates, variations in atmospheric mixing, extent of ion-neutral chemistry, auroral conditions, and~exogenic material influx. To~date, stratospheric column abundances have been measured for a total of 9 neutral molecules and one isotopic D/H (measured in \ce{H2}) with a value of $4.4 \times 10^{-5}$ (Table \ref{Table 4 stratosphere gases}). They are composed of alkanes (\ce{CH4}, \ce{C2H6}), alkenes (\ce{C2H4}), alkynes (\ce{C2H2}, \ce{C3H4}, \ce{C4H2}), and~oxygenated molecules (\ce{CO}, \ce{CO2}, and~\ce{H2O}).
Since no dedicated mission has yet flown to Uranus, new molecular discoveries happen at an irregular pace with long-term windows that depend on telescopic observations \citep{Apestigue2024TheCharacterization}. However, the~benefits of a descending probe into the Uranian atmosphere are undeniable and would increase our understanding of the chemical, radiative, and~optical properties of the atmosphere \citep{Mousis2018ScientificExplorations,Apestigue2024TheCharacterization}. Such an in~situ study would help us constrain the gas and haze composition with the energy deposition distribution in the~atmosphere.

\begin{table}[H] 
\caption{Atmospheric 
 gas phase composition in the stratospheres of Titan, Saturn, and~Uranus. \label{Table 4 stratosphere gases}}
\begin{tabularx}{\textwidth}{lCCCC}
\toprule
\multicolumn{5}{c}{\textbf{Stratosphere}} \\
\midrule
\textbf{Species} & \textbf{Titan}	& \textbf{Saturn}	& \textbf{Uranus} & \textbf{Ref.}\\
\midrule
\ce{CH4}		& 1--2\%			& $4.7\times 10^{-3}$ & 16 ppm & \cite{Niemann2005,Fletcher2009MethaneObservations,Orton2014Mid-InfraredStratosphere} \\
\ce{C2H2}		& $2.97\times 10^{-6}$			& $1\times 10^{-6}$ & 0.25 ppm & \cite{Howett2007MeridionalMeasurements,Guerlet2008EthaneObservations,Sylvestre2015SeasonalObservations,Vuitton2019} \\
\ce{C2H4}		& $1.2\times 10^{-7}$			& $5.9\times 10^{-7}$ & <$2\times 10^{-14}$ & \cite{Hesman2012ElusiveRegion,Orton2014Mid-InfraredStratosphere,Vuitton2019} \\
\ce{C2H6}		& $7.3\times 10^{-6}$			& $1\times 10^{-5}$ & 0.13 ppm & \cite{Howett2007MeridionalMeasurements,Guerlet2008EthaneObservations,Orton2014Mid-InfraredStratosphere,Sylvestre2015SeasonalObservations,Vuitton2019} \\
\ce{C3H4}		& $4.8\times 10^{-9}$			& $1\times 10^{-9}$ & 0.36 ppb & \cite{Guerlet2010,Orton2014Mid-InfraredStratosphere,Vuitton2019} \\
\ce{C4H2}		& $1.12\times 10^{-9}$			& $7\times 10^{-9}$ & 0.13 ppb & \cite{Guerlet2010,Orton2014Mid-InfraredStratosphere,Vuitton2019} \\
\ce{CO2}		& $1.1\times 10^{-8}$			& $4.5\times 10^{-10}$ & 0.08 ppb & \cite{Abbas2013MeasurementsObservations,Orton2014Mid-InfraredStratosphere,Vuitton2019} \\
\ce{CO}		& $4.7\times 10^{-5}$			& $2.5\times 10^{-8}$ & 6 ppb & \cite{Cavalie2009FirstOrigin,Orton2014Mid-InfraredStratosphere,Vuitton2019} \\
\ce{H2O}		& $4.5\times 10^{-10}$			& 1.1 ppb & 3.8 ppb & \cite{Cavalie2019HerschelEnceladus,Moses2020AtmosphericNeptune,Vuitton2019} \\
\ce{D/H} (in \ce{H2}/\ce{C2H2})		& $2.1\times 10^{-4}$			& $2.1\times 10^{-5}$ & $4.4\times 10^{-5}$ & \cite{Coustenis2008DetectionTitan,Pierel2017D/HCIRS,Moses2020AtmosphericNeptune} \\
\bottomrule
\end{tabularx}

\noindent{\footnotesize{Note: Only species shared in common and detected in all three atmospheres are shown here. For~Titan, the~averaged values were taken near the equator by CIRS \citep{Vuitton2019}. Saturn measurements are approximate and derived from CIRS limb data near 400 km \citep{Guerlet2010,Sylvestre2015SeasonalObservations}. Abundances represented here are generally given at or near the mbar pressure level.}}
\end{table}

Multiple photochemical models have been developed, often studied in tandem with Neptune, to~understand the global-averaged hydrocarbon distribution on both \mbox{planets \citep{Atreya1983PhotolysisUranus,Summers1989PhotochemistryUranus,Romani1993MethaneProduction,Lellouch1994TheAtmosphere,Dobrijevic2010,Cavalie2014TheUranus,Orton2014Mid-InfraredStratosphere,Moses2017DustPhotochemistry}}. Generally, they did not however include seasonal and latitude effects on the stratospheric abundance of hydrocarbons. \citep{Moses2018SeasonalNeptune} provided the first one-dimensional (1D) model to incorporate such effects and to track the time- and location-variable influencing the distribution of stratospheric species. This study showed Neptune to have very similar seasonal dynamics to Saturn, while Uranus's were found to be different, primarily due to its 97.8$^{\circ}$ axial tilt and weak vertical transport. The~main hydrocarbons (Table \ref{Table 4 stratosphere gases}) are confined at low altitudes and concentrated at high polar latitudes. The~time constants and photochemical lifetimes of these species (except for \ce{C2H6}) were found to be larger than their loss rates, furthering the stratification of hydrocarbons \citep{Moses2018SeasonalNeptune}. More recently, a~1D seasonal radiative-convective equilibrium model developed by the Laboratoire de Météorologie Dynamique historically used for Jupiter, Saturn, and~exoplanets investigated the origins and evolution of the thermal structure of Uranus and Neptune \citep{Milcareck2024Radiative-convectiveEffects}. This study stressed the importance of knowing the optical properties of haze particles, and~changing optical indices will alter the warming or cooling rates in the atmosphere. In~addition, a~precise \ce{CH4} abundance constraint, with~or without haze, will significantly impact the retrieved temperature profiles. Thus, discrepancies between observations and the simulated thermal structure is likely to hint at variations in the latitudinal stratospheric distribution of \ce{CH4} and haze particles \citep{Milcareck2024Radiative-convectiveEffects}. A~better characterization of these variables will be crucial to understand the extent of Uranus's thermal impact on the atmospheric composition. A~re-analysis of Uranus's energy budget has recently challenged the long-held assumption of a lack of internal heat source \citep{Wang2025InternalUranus}.

\subsection{The Atmosphere of~Titan\label{2.4 section atm of Titan}}

\begin{figure}[H]
\includegraphics[width=14 cm]{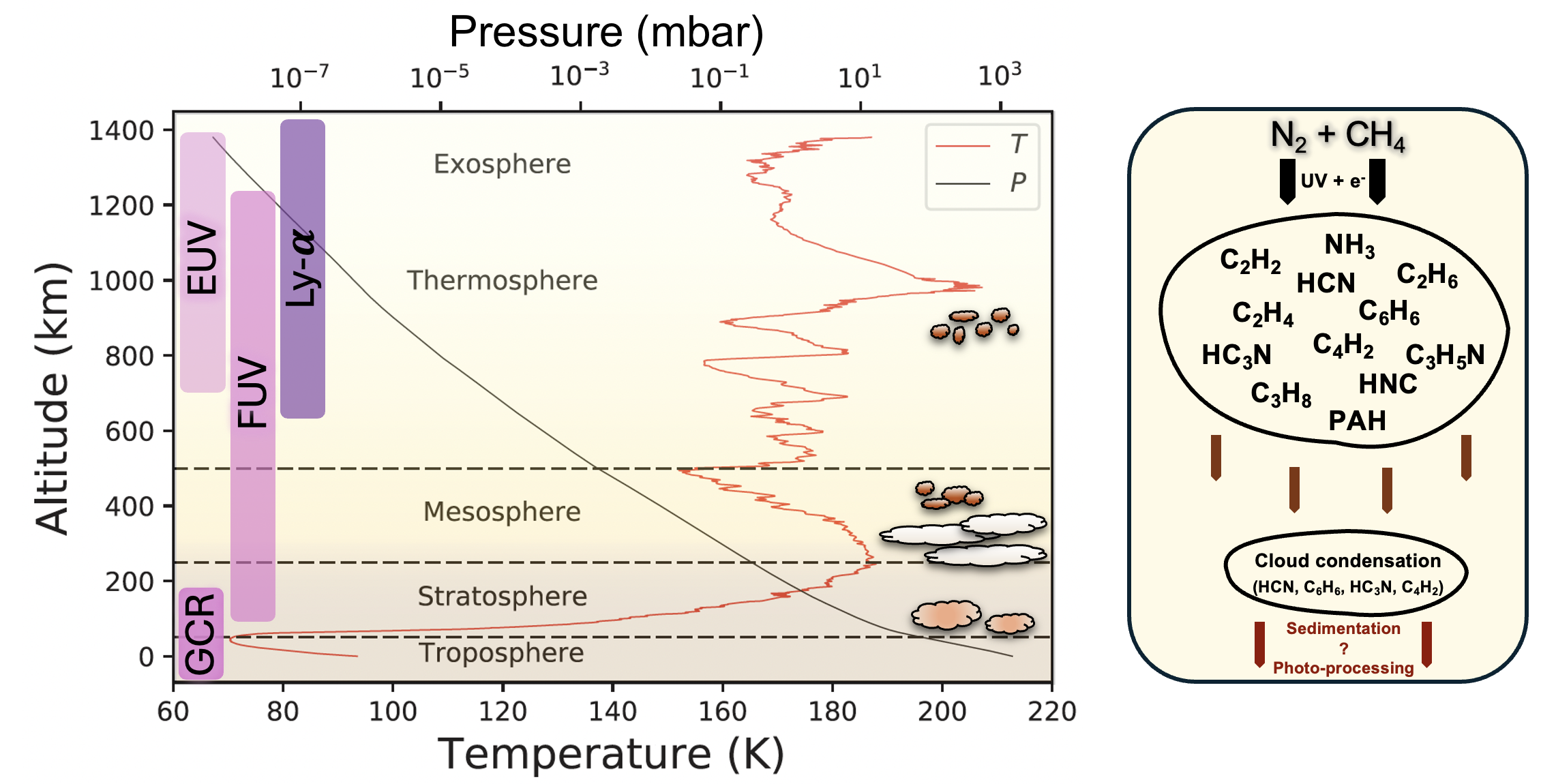}
\caption{Vertical 
 temperature (red) and pressure (black) measurements conducted by Huygens during its descent down to the surface of Titan. Data obtained from the Planetary Data System: Planetary Atmospheres Node. Adapted from \citep{Dubois2018a}. \label{Figure3 Huygens}}
\end{figure}

Titan, the~largest moon of Saturn, is uniquely recognized as the sole known natural satellite to possess a dense atmosphere predominantly composed of molecular nitrogen \ce{N2} ($\sim$98\%) and methane, \ce{CH4} ($\sim$2\%). This reducing atmosphere is stratified into five primary layers---namely, the~troposphere, stratosphere, mesosphere, thermosphere, and~exosphere (Figure \ref{Figure3 Huygens})---each exhibiting planetary-scale dynamical, thermal, chemical, and~seasonal variations. Temperatures are the coldest ($\sim$70 K) at the tropopause. The~formation of the photochemical haze that envelopes Titan is regulated by gas-phase molecular precursors, which are generated in the upper atmosphere through high-altitude \ce{N2} and \ce{CH4} photolysis and radiolysis processes. These precursors consist of hydrocarbon radicals (e.g.,~\ce{CH2}, \ce{CH3}), in~addition to more complex hydrocarbons, nitriles, and~potentially polycyclic aromatic hydrocarbons \citep{Waite2007}. Energetic sources facilitating these high-altitude chemical reactions include solar UV photons, solar X-rays, galactic cosmic rays, Saturn’s magnetospheric energetic electrons, and~the solar wind \citep{Krasnopolsky2014,Plainaki2016}. The~Cassini-Huygens mission, over~a duration spanning 13 years, conducted an extensive study of Titan, taking direct measurements of the composition in neutrals and ions within Titan’s upper atmosphere \citep{Waite2007,Crary2009}. These investigations elucidated the complexity of Titan’s upper atmospheric chemistry, characterized by radicals, neutrals, positive and negative ions, as~well as the early stages of solid haze particle formation. On~numerous occasions, the~measured abundances of specific transient ions such as \ce{CH4+, HCNH+, C2H5+} exceeded predictions from prior models (e.g., \citep{Cui2009}). Conversely, the~concentrations of certain heavy hydrocarbons and nitrogen-bearing neutral molecules, as~determined by the Ion and Neutral Mass Spectrometer (INMS), were found to be lower than anticipated by ion-neutral models (\citep{Cui2009b}, \textit{and references therein
}). These findings emphasized the crucial role of magnetospheric electron precipitation and diurnal solar energy variations in modulating molecular distributions within the upper~atmosphere.

\vspace{-4pt}

Connecting Titan's upper atmosphere dynamics with incoming solar and magnetospheric energy deposition started long before Cassini's arrival at Saturn. Benefiting from earlier Voyager 1's flyby of Titan in 1980 \citep{Broadfoot1981} at a time when Titan was located inside of Saturn's magnetosphere, it became apparent that Titan was at that time in a moon-solar wind 
 configuration akin to Venus \citep{Keller1992}. Well over a decade later, refinements stemming from occultation and plasma experiment measurements led to the expansion of multiple photochemical models \citep{Yung1984,Singhal1986SomeInteraction,Yung1987,Ip1990a,Gan1992,Keller1998} along with models incorporating thermal and suprathermal electrons calculations \citep{Gan1992}. Based on these measurements, it was found that \ce{N2} EUV airglow emission initiated by photoelectron impact dominated over \ce{N2} airglow triggered by magnetospheric electron impact \citep{Gan1992}. Titan's ionosphere operates as a photochemical factory at the interface between incoming solar photon and magnetospheric electron fluxes, and~the underlying first steps of molecular and aerosol growth \citep{Lavvas2008}.

The solar flux reaching Titan is around 14.8 W m$^{-2}$ (Table \ref{Table2 planet parameters}) and the photon flux at Titan ranges from $\sim$$10^6$ ph s$^{-1}$ cm$^{-2}$ nm$^{-1}$ (@60 nm) to 10$^7$ ph s$^{-1}$ cm$^{-2}$ nm$^{-1}$ \mbox{(@110 nm) (\citep{Thuillier2004SolarLevels} and Figure~\ref{Figure XX SOLAR-SSI-HRS})}. As~the primary carrier of VUV energy across the solar system (Region 2, Table~\ref{Table 1 - energy range}), Ly-$\alpha$ (121.6 nm) has the capacity to penetrate into Titan's upper atmosphere where haze particle formation is initiated. VUV radiation reaches a factor of 100 higher in intensity at 121.6 nm than the baseline at 60--110 nm \citep{Thuillier2004SolarLevels}. Longer wavelength UV photons may even reach lower stratospheric levels which continuously exposes the growing aerosols to the UV radiation (\citep{Lavvas2011a} and Figure~\ref{Figure3 Huygens}). Modeling of photon and photoelectron deposition on Titan by \citep{Lavvas2011a} compared with Cassini Plasma Spectrometer Electron Spectrometer (CAPS/ELS) electron flux measurements at 1014 km showed high matching with the observations. Their comparison confirmed the predominant activating role of photon and photoelectron contribution on the dayside compared to the nightside fluxes derived from the contribution of magnetospheric electrons. The~peak photon contribution lies around the ionospheric peak $\sim$1000~km whereas secondary electrons produced from X-rays contribute mostly at lower altitudes (700--900 km). At~the latter altitudes and below the tropopause, GRC radiation is estimated to induce $\sim$10 and $\sim$0.1 ionization events \mbox{cm$^{-3}$ s$^{-1}$}, respectively \citep{Couturier-Tamburelli2014}. At~higher altitude, it is estimated that Ly-$\alpha$ photons account for about 75\% of all photo-dissociated \ce{CH4} \citep{Hebrard2006,Nixon2024}.

Deeper in the atmosphere, shaped by seasonal dynamics, large polar cloud systems have been observed for over 30 years in Titan's stratosphere. Subject to meridional and vertical transport, these stratospheric clouds have been particularly well-studied in polar winter conditions where they contain several ice spectral signatures (see Section~\ref{Section 3}) seen in the mid- and far-infrared \citep{Anderson2018a}. Their formation occur at higher altitudes (>168~km) than predicted by models \citep{Barth2017} and it is likely that they contain ice co-condensates given their overlapping spectral features \citep{Anderson2016,Nna-Mvondo2018}. An~important \ce{HCN}-rich cloud system was even observed at $z=300$ km in the south pole after a substantial cooling event in the polar vortex \citep{DeKok2014,Vinatier2018}. As~a result, climate and circulation models have had to be refined and updated after such observations where they could not account for (i) rapid and dramatic changes in the seasonally defined temperature variations, and~(ii) shortcomings due to lack of theoretical and experimental aerosol microphysics data \citep{Rodriguez2009,Lora2015,Horst2017,Vinatier2018,Barth2020Planetcarma:Atmospheres,Dubois2021,deBatzdeTrenquelleon2025TheStratosphere}. While these discoveries have helped constrain parameters in General Circulation Models (GCM), they have also left many unanswered questions regarding their ice composition and potential for photochemical evolution \citep{Gudipati2015,Anderson2016,Dubois2017b,Couturier-Tamburelli2014,Couturier-Tamburelli2015,Couturier-Tamburelli2018,Mouzay2021}. While the intermediate chemical and physical pathways connecting the formation of aerosols and clouds with the gas phase precursors are still not fully resolved, Titan's chemical catalogue remains one of the most complex in the outer solar~system.

\subsection{Chemical Inventory in~Titan}

According to the most recent photochemical models, Titan's atmosphere is composed of: hydrogen and hydrocarbons, nitrogenated compounds, oxygenated molecules, and~presumed but undetected polycyclic aromatic hydrocarbons (PAH) and reduced phosphorous and sulfur compounds \citep{Wilson2004,Hebrard2012,Hickson2014,Loison2015,Dobrijevic2016,Mukundan2018,Vuitton2019} (See Tables~\ref{Table 3 elemental comp} and \ref{Table 4 stratosphere gases} for summaries of molecules detected, and~\citep{Nixon2024} for an extensive review of Titan's atmospheric composition). The~ionization rate peaks around 1150 km. Employing Chapman layer theory, \citep{Ip1990a} calculated the peak electron density to occur at 1200 km, \citep{Keller1992} at 1175 km, and~\citep{Fox1997} at 1040 km. The~latter study anticipated \ce{HCNH+} to be the dominant ion, utilizing a model comprising over \mbox{60 species} and 600 reactions. Subsequently, Ref. \citep{Keller1998} enhanced this analysis, pinpointing the predominant \textit{m/z} 28 peak as \ce{HCNH+}. Chemical models have helped explain the gas phase chemistry involved in the production of Titan's aerosols (e.g., \citep{Vuitton2019}). Alongside those models, laboratory experiments have helped in vetting the reaction networks through the study of specific channels involving neutral and charged molecular species. Furthermore, photochemical and microphysical models have investigated the dusty nature of Titan’s ionosphere (>900 km in the atmosphere) and characterized the interaction between the aerosols and charged particles \citep{Lavvas2013}. More specifically, by~using a wide array of energy sources and coupling extensive lists of ion-neutral reactions, models have given insights into the first steps linking small hydrocarbons with larger molecules \citep{Dobrijevic2016,Loison2017,Mukundan2018,Vuitton2019}. Ion-molecule reactions are thought to be directly relevant to aerosol growth, and~are controlled by the relative abundances of the two initial neutral main constituents, \ce{N2} and \ce{CH4} \citep{Lavvas2013}. The~dissociation of \ce{CH4} by EUV photons results in the formation of methylene \ce{^1CH2} in its excited state with a dissociation yield of 0.48 for Ly-$\alpha$ photons, and~0.50 in the \mbox{FUV (Reaction 
 (\ref{CH4 photolysis yields CH2 excited})),}
\begin{equation}
\label{CH4 photolysis yields CH2 excited}
\ce{CH4 + h\nu \longrightarrow ^1CH2 + H_2}
\end{equation}

More energetic photons in the EUV ($<$121.6 nm, Table~\ref{Table 1 - energy range}) will preferentially dissociate methane to form the methylene radical \ce{^3CH2} in its ground-state (Reaction (\ref{CH4 photolysis yields CH2 ground state})), while Lyman-$\alpha$ radiation  will also form the methyl radical \ce{CH3} with a dissociation yield of 0.42 (Reaction~(\ref{CH4 photolysis yields CH3}), \citep{Nixon2024}).
\begin{equation}
\label{CH4 photolysis yields CH2 ground state}
\ce{CH4 + h\nu \longrightarrow ^3CH2 + 2H}
\end{equation}
\begin{equation}
\label{CH4 photolysis yields CH3}
\ce{CH4 + h\nu \longrightarrow CH3 + H}
\end{equation}

\begin{figure}[H]
 \includegraphics[width=14 cm]{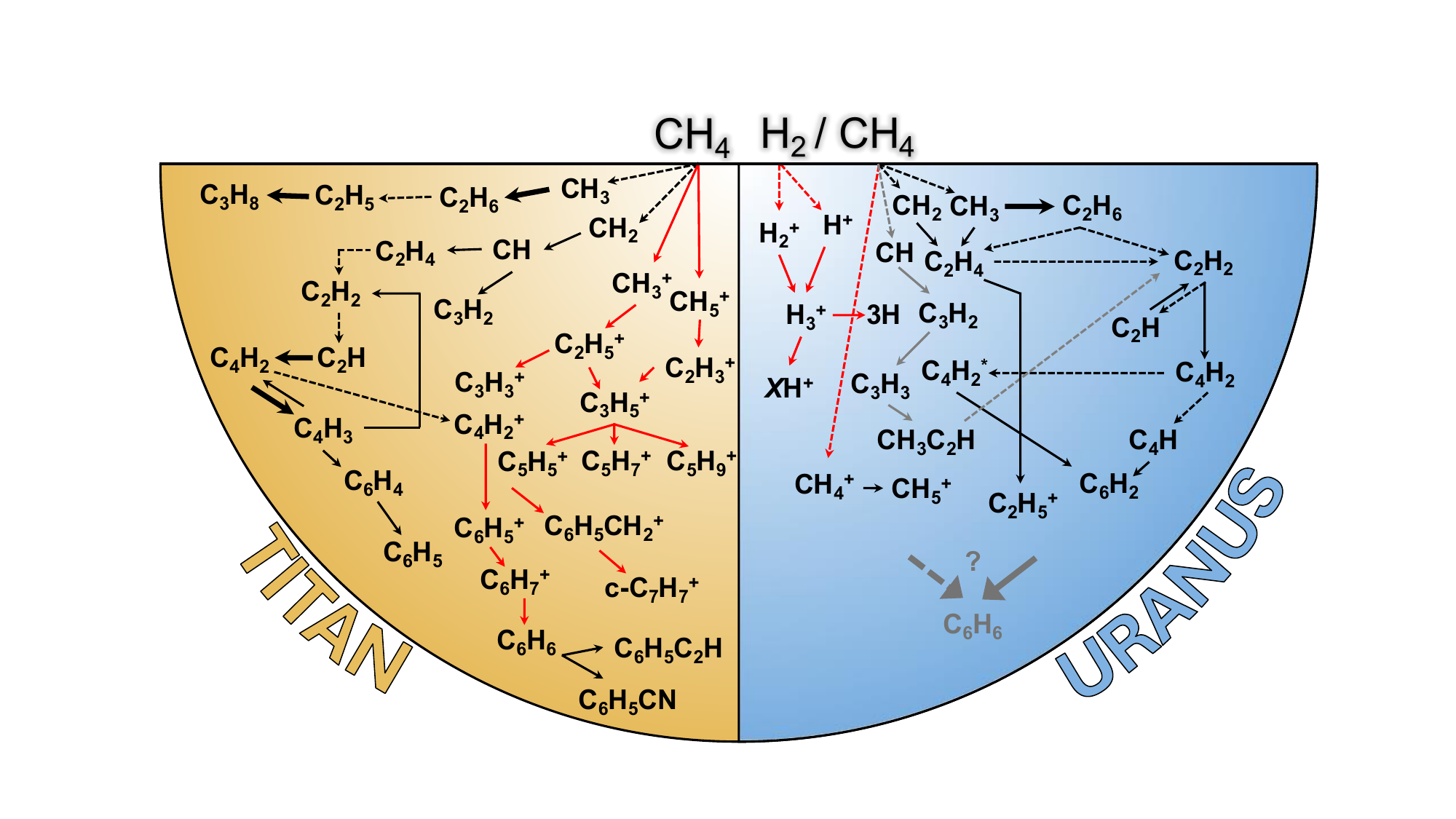}
\caption{Simplified 
 schematic diagram showing the predominant neutral gas phase hydrocarbon formation pathways in the upper atmospheres of Titan and Uranus, based on available combined observations from Cassini, Voyager 2, Spitzer, IRTF, Hubble, and~photochemical \mbox{models \citep{Atreya1975IonosphericNeptune,Atreya1983PhotolysisUranus,Yung1984,Atreya1985PhotochemistryUranus, Chandler1986ThePossibilities,Tyler1986VoyagerSatellites,Atreya1986,Herbert1987The2,Pollack1987NatureHydrocarbons,Waite1987CurrentIonospheres,Summers1989PhotochemistryUranus,Atreya1990UranusNeptune,Atreya1991PhotochemistryMixing,Trafton1993,Moses2005PhotochemistryPlanets,Strobel2005PhotochemistryAtmospheres,Burgdorf2006DetectionSpectroscopy,Orton2014Mid-InfraredStratosphere,Dobrijevic2016,Loison2019,Melin2019TheMorphology,Vuitton2019,Melin2020TheNeptune,Fletcher2021TheUranus,Nixon2024,Pentsak2024,Joshi2025Uranus-hydrogenImages}}. Photolysis reactions are shown with dotted arrows, termolecular reactions in thick arrows, and~all other types of reactions (neutral, proton abstraction, radical-radical, and~radical-molecule reactions) are shown in thin black arrows. Reactions involving ions are shown in red. Note that the reactions in gray leading to the formation of methylacetylene (\ce{CH3C2H}) on Uranus are derived from the photochemical model by \citep{Moses2005PhotochemistryPlanets} developed for the Jovian atmosphere, akin to formation pathways proposed by \citep{Dobrijevic20201DNeptune} to explain the as-of-yet undetected \ce{C6H6} on~Neptune. \label{figure 5 Titan vs Uranus}}
\end{figure}

The formation of these first radicals in Titan's upper atmosphere by FUV, Ly-$\alpha$, and~EUV photons is a crucial starting point towards the formation of the first intermediate, and~then the larger hydrocarbons. Thence begins a long series of cascade of dissociative recombination, proton abstraction, and~polymerization reactions that will ultimately lead to the formation of photochemical aerosols (see \citep{Vuitton2019} and Figure~\ref{figure 5 Titan vs Uranus}). The~absorption of low-energy ($<$10 eV) photons as well as much higher energy EUV, by~radicals is still poorly understood. Long sought for in the Titan community, \ce{CH3} has perplexed photochemical modeling studies for many years due to its non-direct detection. First, because~\ce{CH3} photolysis branching ratios are not well known. Second, \ce{CH3} can be an important source leading to the formation of \ce{^1CH2}. Third, \ce{CH3} production actually expedites \ce{CH4} loss in Titan's atmosphere, where up to 65\% of \ce{CH3} results from \ce{CH4} photolysis with a predicted mole fraction of 10$^{-4}$ at 1000 km, the~highest abundance of any of the small radicals in the atmosphere \citep{Vuitton2019}. As~a result of this complex radical photochemistry, two hydrocarbons are then produced early on (Figure \ref{figure 5 Titan vs Uranus}). First, ethylene \ce{C2H4} forms through the radical-molecule Reaction (\ref{ethylene formation}) and then ethane \ce{C2H6} through the radical-radical Reaction (\ref{ethane formation}) at higher pressure where three-body reactions are permitted, with~a peak production at 800 and \mbox{200 km \citep{Vuitton2019}.}
\begin{equation}
\label{ethylene formation}
\ce{CH + CH_4 \longrightarrow C_2H_4 + H}
\end{equation}
\begin{equation}
\label{ethane formation}
\ce{CH3 + CH3 + M \longrightarrow C_2H_6 + M^\star}
\end{equation}

At 
 these altitudes, FUV photons continue to interact 
 with hydrocarbons such as \ce{C2H2} with low-energy (between 5.5 eV and 12.4 eV) photons which will result in the production of \ce{C2} and \ce{C2H}, two radicals involved in significant methane-depletion and hydrocarbon-formation mechanisms \citep{Nixon2024}. These reactions underscore the contribution of \ce{CH4} photo-degradation by low-energy ($<$20 eV and especially Ly-$\alpha$) photons and the important role radicals play towards the growth of hydrocarbons in a relatively methane-rich environment. This role is particularly important since it also leads to the very molecules that will condense into the lower-altitude clouds and regulate the radiative dynamics of the stratosphere. In~recent observations of the late northern summer, \ce{CH3} was detected in the mid-IR by JWST which showed excellent agreement with photochemical modeling abundance results \citep{Nixon2025TheObservations}, confirming its important role in the neutral~atmosphere.

\vspace{-6pt}
\unskip

\subsection{From Dynamics to Haze Stratification and~Evolution}

As outlined above (Section 
 \ref{Section 2.1}), atmospheric dynamics play a central role in the transport mechanisms influencing photochemical products, haze formation, and~their evolution over time. As~such, comparative planetology among all jovian planets and Titan seems necessary to better constrain the underlying mechanisms at play in these cold atmospheres. Irrespective of the planet, constraining each planet's temperature profile is fundamental to derive other atmospheric properties. 
 For~example, the~refractivity, convection and wind parameters, chemical equilibrium, abundance and condensation of trace gases, cloud formation, precipitation, and~particle size distribution are all processes dependent upon or linked to the local temperature gradient. The~evolution of haze layers (and the condensation levels of precursor trace gases) is of particular interest since its distribution is shaped by the adiabatic regime and temperature profile
~\citep{Guillot1995CondensationPlanets, Leconte2017Condensation-inhibitedNeptune}. With~little data regarding this regime and temperature profile on Uranus, it is thus hard to infer any detailed description of the 
 \ce{CH4} (and other condensates) cloud deck stratification and evolution. Moreover, storms and seasonal variations are also likely to affect downdraft and updraft transport, thus modifying a well-stratified cloud condensation structure
~\citep{Leconte2017Condensation-inhibitedNeptune}. Conversely, a~poorly mixed stratosphere on Uranus may expose certain haze strata to long-timescale radiative influx, thus potentially affecting their composition. Alternatively, subsiding volatile-rich air masses resulting from circulation cell reversal on Titan enabled the formation of cloud condensation at higher altitude than expected 
 \citep{Vinatier2018, Dubois2021}. In~parallel, such transport also allows multiple species (
\ce{C6H6}, \ce{C4H2}, HCN, etc.) to co-condense, resulting in more complex photochemistry than previously thought. Additionally, a~recent re-analysis of solar occultation of Titan prior to the northern spring equinox revealed ``leaking'' 
 \ce{CH4} from the troposphere into the stratosphere, which reinforces the need for better characterization of the thermal fluxes involved 
 \citep{Rannou2021ConvectionStratosphere}. Furthermore, free parameters such as 
 $K_z$ are poorly constrained and in Titan's stratosphere, 
 $K_z$ values used in photochemical models can vary up to a factor of 100 
 \citep{Dobrijevic2016, Dobrijevic2016TheComparison}. In~summary, both Titan and Uranus provide a unique 
 \ce{H2}-rich vs. 
 \ce{N2}-rich portrait to investigate their intrinsic dynamical, thermal, and~chemical behaviors. Ultimately, improved knowledge of the local temperature and compositional conditions where hazes and clouds form can help guide future theoretical and experimental studies that aim to reproduce the compositional and radiative conditions of Uranus.

\section{Photochemistry vs. Radiative Chemistry: Competing Processes and Role in Haze~Formation}
\label{Section 3}
\unskip
\subsection{Fundamental~Processes}\label{sec:fundamentalprocesses}

Photochemical reactions are fundamental to the plurality of chemical inventories in space since they pertain to electronically excited states and are governed by the law of reciprocity (Bunsen-Roscoe law) in that their reactivity is proportional to the fluence, regardless of exposure time. About nine different types of photochemical reactions relevant to planetary and interstellar conditions exist, and~almost twice as many ionizing radiation-based reactions \citep{Petrie2007IonsSpace,Arumainayagam2022Photochemistry,Arumainayagam2022RadiationChemistry}. Among~these, six of the fundamental reactions are listed hereafter:

\begin{itemize}
    \item photodissociation: 
 $\ce{A^* -> B + C}$
    \item quenching: 
 $\ce{A^* + B -> A + B}$
    \item luminescence:
 $\ce{A^* -> A + h\nu}$
    \item photoisomerization:
 $\ce{A^* -> B}$
    \item bimolecular reaction:
 $\ce{A^* + B -> C + D}$
    \item hydrogen 
 abstraction:
 $\ce{A^* + RH -> AH + R}$
\end{itemize}

These reactions occur constantly in planetary atmospheres and serve a fundamental role to initiate gas phase cascade reactions involving radicals and their excited states \citep{Nixon2024}. Photodissociation reactions primarily contribute to the destruction of \ce{CH4} to form the first radicals (see Section~\ref{Section 2}), while proton abstraction is thought to be involved in the loss of \ce{C2H2} to form the first carbon chain anion \ce{C2H-} \citep{Nixon2024}. Other pathways obtained by \citep{Regina2023TheoreticalMedia} applied to the interstellar medium (ISM) 
 have also included bimolecular reactions as an important means of anion growth. Furthermore, infalling \ce{H2O} on Titan is expected to participate in the bimolecular reaction with the excited-state \ce{N(^2D)} atom in the \mbox{ionosphere \citep{Dobrijevic2014}}. Other reactions such as ion-pair formation or sensitization are also included in photochemical models but their involvement not as well understood \citep{Arumainayagam2022Photochemistry}. Note that these channels are driven by photons in the NUV to FUV range. In~the solar system, and~in particular in the outer solar system, low-energy ($<$10 eV) photons may originate isotropically from (inter)stellar background radiation, particularly the interstellar radiation field (ISRF) and scattered Ly-$\alpha$ in the local interplanetary medium (LIPM) resulting from \ce{H2}'s excited Lyman and Werner bands, also called the Solomon process. As~seen \mbox{in \citep{Moses2018SeasonalNeptune}}, accurate modeling of these UV sources are required to understand seasonal effects, particularly at the poles of Titan and Uranus where shadowed conditions can last up to several years in the winter hemispheres. These regions can thus be more favorable to rapid \ce{CH4} photolysis with its dissociation energy threshold of 4.52 eV (Figure \ref{Figure XX SOLAR-SSI-HRS}). Likewise, with~a low-energy dissociation threshold of 4.48 eV, \ce{H2}, a~symmetrical molecule, presents only one vibrational mode as its indirect dissociation occurs through the continuum of the ground electronic state. Notwithstanding limiting cold temperatures, heterogenous chemistry may also come into play where solid grains or polycyclic aromatic hydrocarbons (PAH) may catalyze \ce{H2} formation through chemisorption or physisorption, a~process well-studied in photodissociation and other PAH-enriched regions \citep{Hrodmarsson2025TheReview}.

In addition, radiation chemistry includes pathways pertaining to processes resulting from the interaction between the gas or condensed phases and ionizing radiation above the ionization threshold $\sim$10 eV. A~unique feature emanating from radiation chemistry is the generation of a cascade of low-energy secondary electrons \citep{Pimblott2007ProductionRadiation,Arumainayagam2010Low-energyMatter,Shulenberger2019Electron-InducedIces,Wu2024RoleMolecules}. These electrons's energy have been shown to induce substantial chemical changes on cosmic ice grains, and~have even been proposed to participate in the formation of certain prebiotic molecules \citep{Boyer2016TheAstrochemistry,Stelmach2018SecondaryLife}. In~\citep{Shulenberger2019Electron-InducedIces}, the~authors studied radiolytic mechanisms of degraded \ce{NH3} ices in ISM-like conditions, bombarded with $\sim$1 keV electrons accompanied by low-energy electrons ($\sim$7 eV). The~authors found that non-ionizing radiation of condensed \ce{NH3} (ammonia expected to be the most abundant nitrogen-bearing compound in the ISM, while also present in the gas phase formed in Titan's ionosphere, and~in the condensed state in Uranus's troposphere, Figure~\ref{Figure 2 atm profile Uranus}) was responsible for the formation of hydrazine (\ce{N2H4}) and diazene (\ce{N2H2}) in the condensed phase. It was proposed that a first step involved the dimerization of the \ce{NH2} radical due to $>$1 keV impact, followed by a favorably energetic dissociative electron attachment (DEA) at 6 and 10 eV \citep{Shulenberger2019Electron-InducedIces}. As~surveyed in \citep{Arumainayagam2022RadiationChemistry}, low-energy secondary electrons generated by radiation chemistry can lead to reaction pathways that are otherwise not relevant to photochemistry. Unlike photons, electrons can induce singlet-to-triplet transitions via exchange interactions with electrons \citep{Arumainayagam2010Low-energyMatter}. Additionally, electrons can be transiently captured by molecules, forming temporary negative ions (TNIs) through resonant processes such as shape and core-excited resonances \citep{TP2017Formationsub12/sub,Ryszka2017Low-energyMethyl-pyrrolidine}. These TNIs may undergo dissociative electron attachment (DEA), leading to bond cleavage and the generation of reactive fragments capable of further chemical transformations \citep{Bredehoft2019Electron-inducedPhase,Arumainayagam2022RadiationChemistry}. Through the combination of the yield of secondary low-energy electrons generated by radiation chemistry with UV photons dissociating and exciting molecules, these competing processes contribute to diversifying the molecular inventory necessary for the rapid molecular growth leading to the formation of aerosols in planetary atmospheres. The~following reaction mechanisms are only a handful \citep{Larsson2012,ArumainayagamExtraterrestrialIces}.

\begin{itemize}
    \item Ionization:
 $\ce{AB -> AB+ + e-}$
    \item Ion dissociation:
 $\ce{AB+ -> A+ + B}$
    \item Ion-molecule reaction:
 $\ce{AB+ + BC -> Products}$
    \item Electron attachment:
 $\ce{e- + M -> Products}$
    \item Fluorescence:
 $\ce{AB^* -> AB + h\nu}$
    \item Excimer formation:
 $\ce{A^* + A -> A2^*}$
\end{itemize}

Another process still poorly understood concerns multiphoton and fluorescence chemistry, which was first observed by \citep{Jackson1978MultiphotonPhotochemistry}, resulting from the photodissociation of simple hydrocarbons such as \ce{C2H2} and \ce{C2H4}. In~that study, UV photon (193 nm) absorption alone was not able to explain the fluorescence of free radical photo-products. To~account for this discrepancy, the~authors put forth the proposition of a sequential absorption scheme where molecule \ce{AB^*} absorbs 
 a secondary photon, a~concept revisited in \citep{Jackson1991ImplicationsComets} to understand cometary emission of \ce{C2}. Relevant to Uranus's lower atmosphere, the~multiphoton ionization of \ce{H2S} \citep{Carney1981TheH2S,Achiba1982TheSpectroscopy,Ashfold1982MultiphotonNm} or even PAHs (e.g.,~\citep{Chacko2022MultiphotonInterest}) have been studied in order to understand the ionic fragmentation patterns involved in the UV-VIS region. The~processes involved in radical photochemistry and secondary photolysis still remain to be characterized in the context of planetary atmospheric~chemistry (Table \ref{tab:photo_vs_radchem}).

\begin{table}[H]
\caption{Comparative 
 general characteristics of photochemistry and radiation-induced chemistry in atmospheric and astrochemical~processes. \label{tab:photo_vs_radchem}}

\begin{adjustwidth}{-\extralength}{0cm}
\begin{tabularx}{\fulllength}{lLLL}
\toprule
\textbf{Characteristics} & \textbf{Photochemistry} & \textbf{Radiation-Induced Chemistry} & \textbf{Examples} \\
\midrule
Energy source & Ly-$\alpha$; UV continuum & EUV/X-rays; \mbox{energetic particles} & 100–400 nm; secondary electrons; ions \\
\midrule
Primary effect & Photodissociation; photoionization; \mbox{electronic excitation} & Ionization; radiolysis; dissociative \mbox{electron attachment} & CH$_4$ + $h\nu$ \linebreak $\rightarrow$ CH$_3$ + H \linebreak $\rightarrow$ \ce{H- + CH3+} \\
\midrule
Key products & Radicals; \mbox{small hydrocarbons} & Ions (e.g., CH$_3^+$); complex organics; \mbox{electrons} & \ce{H2+}, \ce{CH3}, \ce{CH3+}, \mbox{\ce{C2H3+}, \ce{C3H4+}} \\
\midrule
Timescales & ns–hours (daylight-driven) & fs–ns (instantaneous, flux-dependent) & H$_2$O → H• + OH• \mbox{(spur reactions)} \\
\midrule
Temperature dependence & Strong (Arrhenius kinetics) & Weak (governed by \mbox{particle flux}) & \ce{CH4 + H} $\rightarrow$ \ce{CH3 + H2} \\
\midrule
Electron transfer & Charge transfer & Ionization cascades; secondary \mbox{electron emission} & \ce{O+ + CH4} \\
\midrule
Quantum effects & Electronic transitions; spin-forbidden pathways & Ro-vibrational excitation; plasmon resonances (ices) & singlet-triplet~absorption \\
\midrule
Observables & Dayglow emissions; \mbox{gas abundances} & Auroral X-rays; Lyman-Werner band emissions; mass spectra \mbox{of ices} & mass spectra, IR-UV~spectra \\
\midrule
Altitude/region & Stratosphere; ionosphere (day side) & Thermosphere; polar auroral zones; \mbox{interstellar ices} & NH$_3$ + $h\nu$ → NH$_2$ + H  \\
\midrule
Desorption yields & Low–moderate (\mbox{UV-photon dependent}) & High (sputtering by \mbox{$>$100 eV} electrons
) & \ce{CO + e-} → \ce{CO_{(ads)}} → \mbox{\ce{CO_(g)}}\\
\midrule
Multiphoton effects & Rare & Dominant (ionization cascades; track formation) & CH$_4$ → CH$_3^+$ + e\textsuperscript{$-$
} (15.6 eV) \\
\bottomrule
\end{tabularx}
\end{adjustwidth}

\noindent{\footnotesize{References: \citep{Menzel1964Electron-impactTungsten, Gans2011PhotolysisSpectrometry, Boduch2015RadiationIces, Hrusak2016StepMethane, Gudipati2013, Shulenberger2019Electron-InducedIces, Arumainayagam2022RadiationChemistry, Arumainayagam2022Photochemistry}.}}
\end{table}

Pathways involving ions are too numerous to include here and the reader is referred to~\cite{Moses2018SeasonalNeptune,Vuitton2019,Nixon2024} for an updated in-depth overview of known reaction schemes for Titan and Uranus. Instead, a~focus hereafter will be placed on the competition between photochemical vs. radiative processes (impacting excited and dissociative ion chemistry, respectively) since ion reactivity remains a crucial component to investigate to understand haze formation in the solar system, interplanetary, and~interstellar ices \citep{Larsson2012,Arumainayagam2022Photochemistry,Arumainayagam2022RadiationChemistry,Wu2024RoleMolecules}. Numerous studies have investigated the photochemical effects induced by Ly-$\alpha$ on interstellar matter which represents a good benchmark for heterogeneous chemistry conducive to the formation of organic molecules in the solar system \citep{Boyer2016TheAstrochemistry,Boyer2019Low-energyApplications,ArumainayagamExtraterrestrialIces,Arumainayagam2021ExtraterrestrialApplications,Wu2024RoleMolecules}. Furthermore, FUV, EUV, and~even NUV radiation has been shown to induce chemistry in the condensed state in laboratory Titan ice and aerosol analogues \citep{Gudipati2013,Couturier-Tamburelli2014,Couturier-Tamburelli2015,Couturier-Tamburelli2018,Mouzay2021}. In~Titan's upper atmosphere, photochemically produced haze will indeed interact with these UV photons, thus modifying their composition and the ice CCN down to mesospheric and stratospheric altitudes. The~extent of this $h\nu$-induced chemistry remains far from being fully characterized, and~the task becomes even worse for Uranus. Here, we will present a summary of studies that have probed Titan's and/or Uranus's photochemical processes vs. those that have investigated radiation chemistry at energies $<50$ eV utilizing (\textit{primus
}) Cassini/Voyager or ground-based observations, (\textit{secundus
}) experimental analyses simulating Titan-like low-temperature reactivity in the condensed state initiated by low-energy ($<$20 eV) photons and electrons, (\textit{tertius
}) advances in the theoretical study of branching ratios and their relevance in the VUV, and~(\textit{quartus
}) considerations of the relatively poorly understood role of dication species and their reactivity through vertical ionization~processes.


\subsection{Observational Considerations: From Low-Mass to Intermediate-Mass~Molecules}
\subsubsection{Low-Mass~Species}

\begin{figure}[H]
\includegraphics[width=13.6 cm]{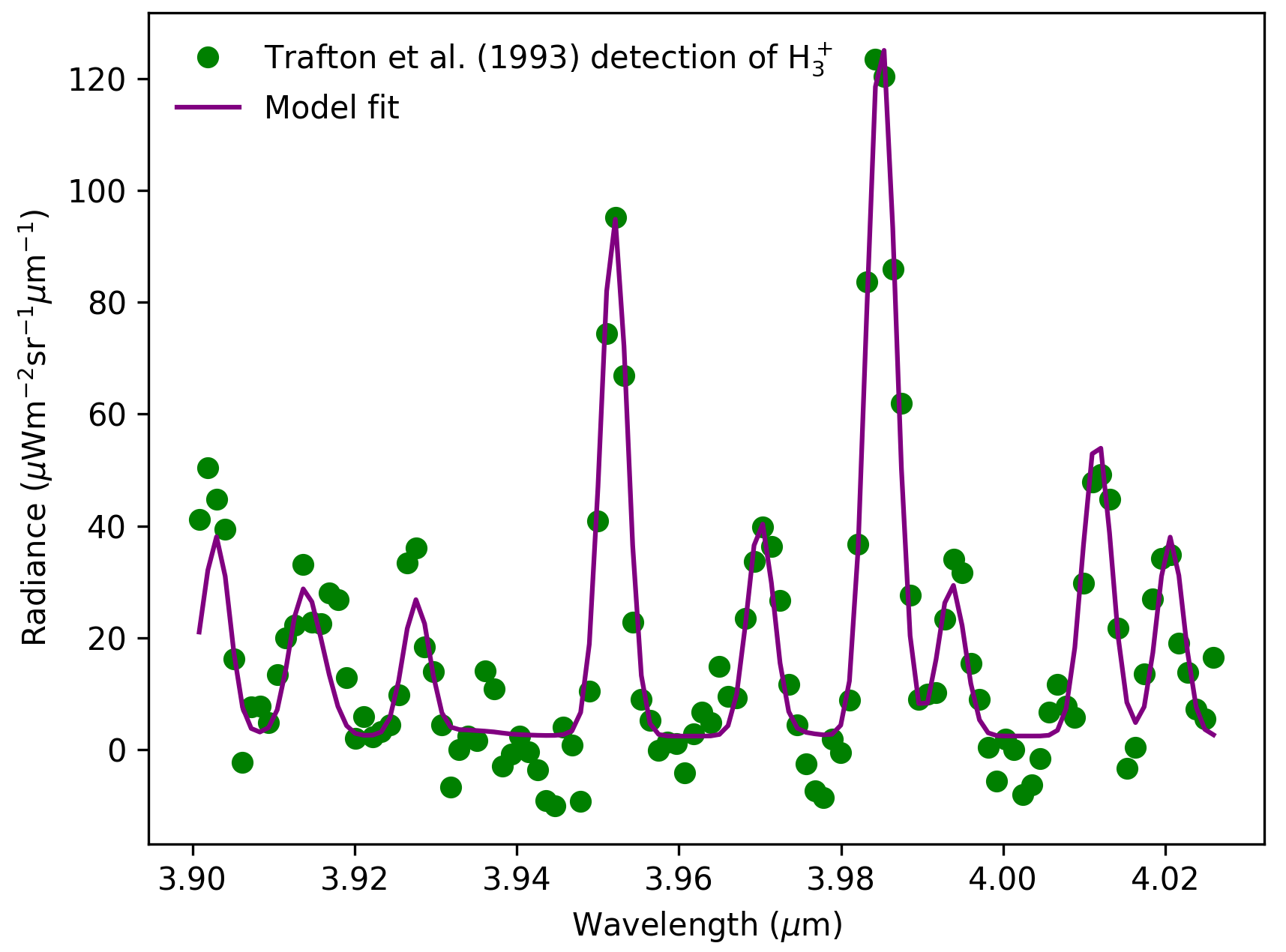}
\caption{Observed spectrum of \ce{H3+} (green dots) detected by \citep{Trafton1993} between 3.89 and 4.03 $\upmu$m, a~range which includes the region with multiple lines of the \textit{Q} branch of \ce{H3+}, with~the most intense \textit{Q}(3) band at 3.985 $\upmu$m \citep{Melin2019TheMorphology}. A~model fit using the \texttt{h3ppy
} open source package computed with a rotational temp 740 K is shown in purple. Here, the~model retrieves an \ce{H3+} column density of $4.42 \times 10^{14}$ m$^2$ (\url{https://github.com/henrikmelin/h3ppy}, accessed on 21 May 2025
). \label{Figure 6}}
\end{figure}

As part of this discussion and the species discussed below, we will follow the simplified classification from \citep{Yung1999PhotochemistryAtmospheres,Plane2001PhotochemistryAtmospheres} separating low Z and high Z elements found in Jovian planets, resulting from photochemistry and radiation chemistry processes. Low Z elements revolve around \ce{H} and \ce{He}, while high Z elements concern any species with a mass higher than \ce{He}. A~focus will be placed primarily on the role of H on Titan and Uranus, and~\ce{H3+} on Uranus. About 20 years separate the first detection of \ce{H2} on Uranus \citep{Herzberg1952SpectroscopicNeptune} and on Titan \citep{Trafton1972b}. \mbox{Two decades} later, \ce{H3+} was detected on Uranus, serving as an important temperature proxy in the upper atmosphere from its emission intensity-dependence on solar \mbox{activity \citep{Miller2000TheAtmospheres,Trafton1993}}. Figure~\ref{Figure 6} shows the observed spectrum of \ce{H3+} and the intense \textit{Q}(3) band centered at \mbox{3.985 $\upmu$m} with the open source model fit using the \texttt{h3ppy
} package (version 0.6.1). The~chemistry of \ce{H3+} on Uranus mainly involves daytime and nighttime processes. EUV photons (>10 eV) on the dayside of Uranus ionize \ce{H2} which produces \ce{H2+} and a free electron (Reaction (\ref{H2 ionization})). \ce{H2+} then reacts with the background \ce{H2} to produce \ce{H3+} through the exothermic Reaction (\ref{H3+ formation}). Recently, other channels have been proposed to explain \ce{H3+} formation, requiring impact ionization of \ce{C2H6} by higher energy electrons (300 eV) \citep{Zhang2020FormationImpact} and from doubly ionized cyclopropane \ce{C3H6} \citep{Kwon2023WhatCyclopropane}. The~atomic yield of hydrogen through these channels is an important consideration since, as~pointed out in \citep{Summers1989PhotochemistryUranus}, (i) H loss by chemistry (\mbox{Reactions (\ref{H loss 1}) and (\ref{H loss 2})}) influences the recycling and abundance of \ce{CH4}, and~\mbox{(ii) controls} the production of polyacetylenes such as \ce{C4H2} from competing quenching of metastable excited \ce{C4H2^{**}} by \ce{H2} and bimolecular reaction pathways. Note that \ce{H3+} was recently detected for the first time at Neptune with JWST \citep{Melin2025DiscoveryJWST}.
\begin{equation}
\label{H2 ionization}
\ce{H2 + h\nu \longrightarrow H2+ + e-}
\end{equation}
\begin{equation}
\label{H3+ formation}
\ce{H2+ + H2 \longrightarrow H3+ + H}
\end{equation}
\begin{equation}
\label{H loss 1}
\ce{H + C4H2 + M \longrightarrow C4H3 + M}
\end{equation}
\begin{equation}
\label{H loss 2}
\ce{H + C4H3 \longrightarrow C4H2 + H2}
\end{equation}

\ce{C4H2} being a strong absorbent of UV light, it is likely to be excited into the metastable state \ce{C4H2^{**}} between 180 and 260 nm, a~process more likely to occur on Uranus than in the nitrogen-rich atmosphere of Titan \citep{Summers1989PhotochemistryUranus}. These reactions were later taken into account in \citep{Atreya1990UranusNeptune}, and~the later detection of \ce{C4H2} by \citep{Burgdorf2006DetectionSpectroscopy} raises the question of what the extent of these small hydrocarbons on Uranus is and the competing roles between photochemistry, radiation chemistry, and~condensation processes which might affect the abundances of polyacetylenes in the atmosphere. Furthermore, as~shown in stratospheric photochemical modeling \citep{Moses2018SeasonalNeptune}, the~latitudinal distribution of hydrocarbons is strongly dependent on seasonal effects (i.e.,~solar radiation), and~thus on photochemical efficiency. On~Titan, the~chemistry of low Z elements is relatively less influential than that on Uranus, since \ce{H2} concentrations on Titan are much lower ($\sim$0.4\% \citep{Waite2005}). Still, atomic hydrogen is thought to originate mainly from \ce{CH4} photolysis (other subsurface sources may also exist \citep{Lorenz2019}), although~its exact vertical profile and source/sink equilibrium are still debated (\citep{Nixon2024}, \textit{and references therein
}).

\unskip 

\subsubsection{Higher-Mass~Species}

Higher Z-mass species (\textit{m/z} $>$ He) between Titan and Uranus share one common chemical entry point: the photo-destruction of \ce{CH4} (Figure \ref{figure 5 Titan vs Uranus}). As~discussed above, \ce{C4H2} with a mass of 50 amu remains the highest mass of a molecule detected on Uranus, while negatively charged particles with masses of over $>$10,000 amu are the largest detected on Titan (albeit with no exact molecular identification to date). If, on~Uranus, intermediate mass molecules (\ce{C2H2}, \ce{C2H6}, \ce{C4H2}) do condense out near their production altitude \citep{Atreya1990UranusNeptune}, UV-driven solid-state chemistry becomes a possibility, although~this is at present a field almost completely unexplored. As~pointed out in \citep{Loison2019}, photolytic destruction branching ratios of \ce{C2H2} and formation kinetics of \ce{C2H6} from \ce{^1CH2} are also not well constrained and model-sensitive, which can result in significant modeled abundance variations. The~formation of PAHs is also an unresolved question, as~they have only been indirectly identified on Titan \citep{Lopez-Purtas2013}, while \ce{C6H6} alone has so far not been detected on Uranus. Recent experimental work in this field has shown promising results (see Section~\ref{section 3.3}). In~order to broaden our understanding of Uranus's molecular distributions, it may be advantageous to focus future research efforts on the reactivity of \ce{C2H2} with other primary hydrocarbons. Improving our understanding of the chemistry of this compound, a~precursor to polyynes, cyanopolyynes, and~potentially PAHs, may facilitate the future detection of novel species \citep{Pentsak2024}.

As seen in Section~\ref{2.4 section atm of Titan} and Figure~\ref{figure 5 Titan vs Uranus}, the~net end result of \ce{CH4} photolysis due to UV photons results in high Z hydrocarbons (Table \ref{Table 4 stratosphere gases}). With~Titan and Uranus having reducing atmospheres, high-altitude haze formation benefits from the supply of EUV radiation which in turn heats the haze layers and creates the observed high-altitude temperature inversions (see Figure~\ref{Figure3 Huygens}). These mechanisms have been well characterized on Titan and remain poorly known on Uranus \citep{Mousis2018ScientificExplorations}. In~spite of the tropopause cold-trap, condensing species such as \ce{CH4} manage to rise into the stratosphere under unknown \mbox{mechanisms \citep{Moses2020AtmosphericNeptune}}, possibly due to moist convection. This transport brings \ce{CH4} to higher altitudes (an effect even more pronounced on Neptune due to stronger mixing) whereby VUV photons begin photodissociation and photoionization \citep{Moses2020AtmosphericNeptune}. The~most abundant photolytic product, \ce{C2H6} is important on Titan for several reasons. First, it is one of the first photochemically produced hydrocarbons and reaches $\sim$10\textsuperscript{$-$6} in the stratosphere (Table \ref{Table 4 stratosphere gases}) with detections of ice spectral signatures thought to correspond to ethane ice condensation above the tropopause over the north pole \citep{Coustenis1999,Griffith2006}. Second, its formation involves ion-neutral reaction coupling associated with methane loss via Reaction (\ref{ethane formation}). Also, once formed, neutral ethane is likely to react with C2 cations resulting in the formation of C3 and C4 ions \citep{Vuitton2019}, thus accelerating molecular growth. In~the EUV/FUV wavelength range, \ce{C2H6} destruction mainly leads to the formation of ethylene \ce{C2H4} and two hydrogen atoms, while at longer wavelengths ($>$140~nm) \ce{C2H4} is formed along with \ce{H2}, with~branching ratios of \mbox{0.30 and} 0.12, respectively (see 
Section~\ref{section branching ratios}). This is an important step since not only is ethane an important intermediate in the photochemical production of atomic and molecular hydrogen on Titan (although the majority of hydrogen is formed thanks to \ce{CH4} photolysis), but~also because DEA of \ce{H2} may thus form the \ce{H-} anion where low-energy (4 eV) trapped electrons reside in a resonant state within \ce{H2} \citep{Drexel2001DissociativeHydrogen}. Molecular hydrogen is also efficiently produced through the photodissociation of ethylene \ce{C2H4} with UV photons between 118 and 175 nm \citep{Gans2013,Hrodmarsson2023PhotodissociationDatabase} through the following reaction pathway:
\begin{equation}
\label{C2H4 photodiss1}
\ce{C2H4 + h\nu \longrightarrow C2H2 + H2}
\end{equation}

This constitutes an important reaction since (i) ethylene (first discovered by \citep{Gillett1975FurtherTitan}) is the smallest alkene (i.e.,~the two C atoms are linked by a double bond) and the only photochemical hydrocarbon to never condense at Titan's tropopause, and~(ii) photolysis of \ce{C2H4} incorporates a $\pi\rightarrow\pi^*$ excitation where absorbed low-energy photons ($<$10 eV) lead to an electronically excited state of the molecule. In~this excited state, ethylene may go through photoisomerization or proton abstraction reactions, Furthermore, ion-neutral reactions play here a key role for molecular growth by producing some of the first C3 (Reaction (\ref{C2H4 loss})) and then C5 (Reaction (\ref{C2H4 loss2})) ions, what has been described by \citep{Nixon2024} as a ``stepping stone from methane to higher hydrocarbons''. The reactivity of \ce{C2H4} with \ce{N(^2D)}, the~first electronically excited state of atomic nitrogen produced by VUV photons, has been studied on Titan and plays a key role in the loss of \ce{N(^2D)} as well as the formation of nitrogenated products \citep{Hickson2020ATemperature}. Overall, photochemistry is thus a fundamental mechanism to rapidly convert small molecules into larger organics.
\begin{equation}
\label{C2H4 loss}
\ce{C2H4 + C2H5+ \longrightarrow CH4 + C3H5+}
\end{equation}
\begin{equation}
\label{C2H4 loss2}
\ce{C2H4 + C3H5+ \longrightarrow H2 + C5H7+}
\end{equation}

\subsubsection{Polycyclic Aromatic Hydrocarbons: Agents of Haze Growth?}

On Titan, the~composition of gas phase products with masses $>$ 78 amu, corresponding to \ce{C6H6}, is still largely unknown. Data in this molecular mass regime (heavy ions and neutrals) is austere and can be summarized in three broad categories, though~with no exact molecular formulae. First, aliphatic compounds that comprise \ce{C2H2} polymers, nitrogenated polymers, and~aliphatic copolymers \citep{Crary2009}. These broad mass groups (possibly up to C14) have only been detected in their positively charged form by Cassini, at~altitudes where EUV radiation peaks in the ionosphere ($\sim$1000 km). Second, PAHs and nitrogenated-PAHs have been tentatively detected, with~naphthalene (\ce{C10H8+}) and anthracene (\ce{C14H10+}) being candidate cations \citep{Crary2009,Hrodmarsson2025TheReview}. Ion-neutral pathways are likely dominant processes for their formation. In~the gas phase, VUV photons are generally needed to ionize PAHs through radiation chemistry. Recently, Ref. \citep{Lignell2021Visible-lightEnergy} investigated the low-temperature visible-light photoionization of PAHs trapped in crystalline ice. The~authors found that trapped PAHs in low-temperature water ice had their ionization threshold lowered by 4.4 eV with respect to their neutral gas phase counterparts. This opens new venues towards the characterization of potential PAHs in the atmospheres of Titan and Uranus. Third, negatively charged compounds (anions) have long remained a core question related to the gas composition of the upper atmosphere of Titan since their discovery by \citep{Coates2007}. More details on them are given in Section~\ref{Section 4}.

To date, neither benzene or PAHs have not been detected on Uranus (or Neptune). However, a~point should be made here to address the photochemistry of \ce{C6H6} and PAHs of relevance to Uranus, through comparative planetology with Jupiter and Saturn. A~search for \ce{C6H6} was conducted on all gas giants by \citep{Bezard2001BenzenePlanetsb} using the Infrared Space Observatory. While abundance values were determined for Jupiter and Saturn, only upper limits were provided for Uranus and Neptune. Early photochemical modeling by \citep{Moses2005PhotochemistryPlanets} containing only neutral-neutral reactions and Ly-$\alpha$ radiation originating from the LIPM did not predict the formation of \ce{C6H6} on the two ice giants, and~acknowledged the need for accurate \ce{C6H6} absorption cross-sections, branching ratios at low temperature and low pressures. Loss mechanisms part of their model included \ce{C6H6} destruction by photolysis to form either a phenyl group (\ce{C6H5}) or \ce{C6H4}. Another important assumption, based on Jovian photochemistry, was that \ce{C6H6} in an excited ``hot'' state stabilizes through collisions once transported down to Jupiter's lower stratosphere \citep{Moses2005PhotochemistryPlanets}, although~the authors also note the possibility for benzene to be produced by ion-neutral reactions in aurorae-rich regions. Putative stabilization of benzene may thus prevent the molecule from dissociating \citep{Moses2005PhotochemistryPlanets}. The~non-detection of benzene on Uranus (or Neptune), however, has renewed interest in incorporating ion chemistry to models, particularly when using Titan as a baseline \citep{Loison2019,Dobrijevic20201DNeptune}. In~\citep{Dobrijevic20201DNeptune}, coupling ion-neutral pathways brought predicted mole fractions of benzene from $\sim$10\textsuperscript{$-$13} (neutral reactions only) to $\sim$10\textsuperscript{$-$9}--10\textsuperscript{$-$10} (ion-neutral reactions) between $1~\upmu$bar and 1 mbar in Neptune's stratosphere. We can expect there to be similar reactions occurring in Uranus's atmosphere, although~important caveat variables exist such as (i) temperature-dependent photodissocative branching ratios, (ii) electron dissociation recombination rates, (iii) cloud microphysics and equilibrium vapor pressure measurements at low temperature, and~(iv) vertical transport \citep{Dobrijevic20201DNeptune,Dubois2019IAUS350,Dubois2021}. Furthermore, photochemical reaction pathways, either omitted from models or whose branching ratios are unknown, involving \ce{C2H2} or its derivatives, remain to be explored \citep{Pentsak2024}. As~pointed out in \citep{Vuitton2008}, laboratory measurements of electron recombination rates of \ce{C6H7+} by low-energy electrons ($<$1 eV, Reaction (\ref{C6H6 form1})) with a rate constant of ($\alpha=2.4\times10^{-6}$ cm$^3$ s$^{-1}$ valid for electrons temperatures between 300 and 800 K) and their products are key yet unknown parameters needed to understand the formation pathways leading to benzene. Other pathways such as the trimerization Reaction (\ref{C6H6 form2}) may also be relevant to surface or tropospheric conditions on Titan due to GCR radiation \citep{Pentsak2024}. Once \ce{C6H6} is formed, PAH formation may also be possible through the ethynyl addition mechanism (Reaction (\ref{C6H6/PAH form1})) for which both Titan and Uranus may provide suitable low-temperature, hydrogen-rich environments \citep{Mebel2008PhotoinducedAdditions,Pentsak2024}.

\begin{equation}
\label{C6H6 form1}
\ce{C6H7+ + e- \longrightarrow C6H6 + H}
\end{equation}
\begin{equation}
\label{C6H6 form2}
\ce{3C2H2 + GCR \longrightarrow C6H6}
\end{equation}
\begin{equation}
\label{C6H6/PAH form1}
\ce{C6H6 + C2H \longrightarrow C6H5C2H + H}
\end{equation}

Finally, catalytic reactions on the surface of interplanetary grains with the interaction of \ce{C2H2} with \ce{SiC} has been shown to be an efficient mechanism to produce cyclic and/or prebiotic molecules (see \citep{Pentsak2024} for a detailed review on the matter). Delivery of SiC grains (found in meteorites and comets) and subsequent chemistry with \ce{C2H2} may lead to \ce{C6H6} formation, while silicon dicarbide \ce{c-SiC2} (found in the ISM) may be produced by the UV photolysis of \ce{c-SiC2H2} (Reaction (\ref{catalytic Titan}), \citep{Parker2013OnMedium,Pentsak2024}). The~kinetics of these reactions, and~thus their capacity to act as efficient catalyzers for larger organics formation, remains an open field of study, one of interest for the cold outer solar system. More recently, quantum chemistry calculations explored gas phase cyanobenzene formation routes catalyzed by the \ce{NCN-} anion \citep{Esposito2025AIntermediates}.
\begin{equation}
\label{catalytic Titan}
\ce{c-SiC2H2 + h\nu \longrightarrow c-SiC2 + H2}
\end{equation}




\subsection{Laboratory Experiments: Simulating Atmospheric~Chemistry}
\label{section 3.3}
From the first seminal set of studies designed to mimick the UV- and magnetospheric-field of radiation of an outer solar system body (Titan), laboratory experiments have unveiled many processes involved in the formation of laboratory analogues of planetary aerosols called \textit{tholins
} \citep{Cable2012,Horst2017}. These studies have used multiple sources of energy to simulate either the UV photon radiation reaching the upper atmosphere of Titan, or~more energetic Ly-$\alpha$ or GCR radiation. In~the former case, electrical plasma discharges have been widely used as a realistic analogue of solar radiation-induced chemistry as impact from inelastic electron collisions will dissociate the carrier gases \ce{CH4} and \ce{N2}, and~their overall electron energy distribution functions resembles that of the solar spectrum \citep{Thuillier2004SolarLevels}. The~electrons deposited on \ce{N2}-based gas mixtures thus lose their energy through inelastic collisions and are able to initiate the chemistry between \ce{N2}, \ce{CH4}, and~any other gas present in the mixture. To~reproduce the more intense radiation such as Ly-$\alpha$ photons, GCRs, or~higher-energy photons, dedicated microwave plasma sources or synchrotron facilities (\mbox{for energies < 200 nm}) have been utilized. A~panorama of these facilities used to simulate Titan's atmospheric chemistry
, from~those studying gas phase to condensed-phase chemistry, is shown in Table~\ref{Table 5}. The~reader is referred to \citep{Cable2012} for a historical overview of experimental simulations of \textit{tholin
} formation to simulate Titan's~atmosphere.

\begin{table}[H]
\caption{Panorama of past and current experimental facilities enabling the chemical study in the laboratory of simulating various typical energy sources reaching Titan's atmosphere. The~broad objectives for each experimental setup are also annotated. SE correspond to low-energy secondary~electrons.\label{Table 5}}

\begin{adjustwidth}{-\extralength}{0cm}
\begin{tabularx}{\fulllength}{m{3cm}<{\raggedright}m{5cm}<{\raggedright}cC}
\toprule 
\textbf{Category $^1$} & \textbf{Main Processes} & \textbf{Energy Source} & \textbf{References} \\
\midrule 
\multirow{2}{*}{Gas 
 phase} 
  & \makecell[l]{Ionization, dissociation,\\radical chemistry} 
  & Plasma discharges  
  & \citep{Sanchez1966,Gupta1981,Khare1981,Scattergood1989,Thompson1991,McDonald1994,Coll1995,Navarro-Gonzalez1997CoronaTroposphere,Ramirez2002,Imanaka2004,Ramirez2005,Szopa2006,Ricketts2011,Carrasco2012,Horst2013haze,Sciamma-OBrien2014,CunhaMiranda2016,Dubois2019a,Dubois2020,Perrin2021,He2022TitanExperiments} \\

  & \makecell[l]{Photolysis, radical, \\excitation, SE}         
  & FUV--Ly-$\alpha$--EUV      
  & \citep{Dodonova1966,Sagan1971,Chang1979,Ferris2005,Vuitton2006,Imanaka2009,Horst2013haze,Carrasco2013,Tigrine2016,Gautier2017InfluenceAnalogues,Carrasco2018,Berry2018,Bourgalais2019,Bourgalais2021AromaticChemistry,Carrasco2022} \\

Tholins/ice 
  & \makecell[l]{Condensation, solid-state\\photochemistry, SE}        
  & FUV/VUV       
  & \citep{Curtis2008,Gudipati2013,Couturier-Tamburelli2015,Couturier-Tamburelli2014,Sciamma-OBrien2017,Couturier-Tamburelli2018,Couturier-Tamburelli2018BehaviourTitan,Fleury2019,Selliez,Lignell2021Visible-lightEnergy,Mouzay2020,Mouzay2021,Couturier-Tamburelli2025PhotochemistryAtmosphereb} \\
\midrule
 Synchrotron & \makecell[l]{Ionization, dissociation,\\excitation, SE} & \makecell[c]{EUV-VUV\\Target wavelength} & \citep{Imanaka2007,Imanaka2009,Thissen2009,Peng2013,Carrasco2013,Carrasco2018,Tigrine2018}\\
\bottomrule
\end{tabularx}
\end{adjustwidth}
\noindent{\footnotesize{$^1$ Note: experimental needs for Uranus using similar techniques are listed below.}}
\end{table}

Currently, laboratory simulations of Uranus's atmosphere are severely limited. As~discussed previously, multiple parameters needed to better understand our comprehension of photochemical haze formation under low-energy (<50 eV) particles are lacking from state-of-the-art photochemical models. Such parameters and uncertainties can be corroborated by experimental simulations using analogous Titan techniques as a basis to probe the chemistry in 
 \ce{H2}-based environments. 
 Laboratory experiments such as those listed in Table
~\ref{Table 5} are in an opportune time to address these knowledge gaps, whether in determining accurate reaction kinetics, branching ratios, or~in studying the photochemical evolution of organic and sulfuric Uranus-relevant ices. Following the themes outlined in 
 \citep{Mousis2018ScientificExplorations,Moses2020AtmosphericNeptune}, more details and future perspectives are given below and 
 in Section~\ref{Section 5 conclusions}.

\begin{itemize}
    \item  \textit{Atmospheric chemistry science:
} Multiple complementary laboratory experiments (Table 
 \ref{Table 5}) utilizing different sources of energy (plasmas, UV lamps, high-energy synchrotron beamlines) are substantial to probe specific chemical and photoionizing processes. In~particular, questions surrounding the role of ion-molecule chemistry and haze growth on Titan have significantly benefitted from laboratory studies. Coupled with \textit{in~situ} or \textit{ex~situ} analyses such as high-resolution mass spectrometry, IR spectroscopy, electron microscopy, secondary ion mass spectrometry, X-ray photoemission spectroscopy, atmospheric-pressure photoionization, just to name a few, future measurements would provide much insights into the chemical composition of Uranian 
 \textit{tholins
} and gas phase precursors. Furthermore, laboratory characterizations of photochemical products would help support \textit{in~situ} measurements by a future UOP and help quantify these precursors resulting from the photodissociation of 
 \ce{CH4}, \ce{NH3}, etc.
    \item  \textit{Cloud science:
} Laboratory measurements of the physical and optical properties of any Uranian laboratory-produced aerosols would directly provide valuable information to interpret cloud observations and the modeled scattering, nucleating, and~size properties of the CCN. Their properties would then help address the role and interaction of clouds with condensable species. Moreover, studies of the photochemical evolution under low-energy (as well as of much higher-energy) photon irradiation remains critically unexplored.

    \item \textit{Chemical kinetics \& thermodynamics:
} As outlined below, kinetic rates, branching ratios, and~absorption cross-sections are fundamental properties that are needed to solve model degeneracies and inaccurate abundance retrievals (Table \ref{tab:branching-ratios-neutrals}). Future theoretical calculations combined with experimental measurements are much needed.

\end{itemize}

\subsection{Condensed~Phase}

Ice cloud spectral signatures have been detected on both Titan and Uranus (see \mbox{Section~\ref{Section 2}}). The~clouds in these atmospheres are tropospheric/stratospheric and can be at similar altitudes to the photochemical hazes \citep{Horst2017,Anderson2018,Fletcher2021TheUranus}. Over~more than a decade, multiple studies have investigated the low-temperature reactivity of Titan ice cloud analogues under UV photon irradiation. These studies have also measurement the fundamental properties of these Titan-relevant ices/\textit{tholins 
} with no energy input, e.g.,~IR absorption \mbox{spectroscopic \citep{Anderson2018,Nna-Mvondo2018}} and their optical properties \citep{Sciamma-OBrien2023FirstAerosols}, and~vapor pressures \citep{Dubois2021,Hudson2022BenzeneResults}, as~well as UV photon and synchrotron light irradiation (see Table~\ref{Table 5}). Reproducing stratospheric ice photochemistry has had to adapt to the ever-evolving knowledge of Titan's stratospheric composition and seasonal variations. Nonetheless, ice mixtures have typically been formed from any of the following molecules \ce{CH4}, \ce{C2H2}, \ce{C4H2}, \ce{C6H6}, \ce{HCN} (Table \ref{Table 5}). The~overall objective of these studies can be summarized in two broad categories: (i) reproducing spectral features seen by Cassini's Composite Infrared Spectrometer (CIRS) in the mid- and far-IR, and~(ii) measuring the spectroscopic and chemical impacts of FUV photon irradiation on these ices. In~the latter case, few laboratory apparatus have explored the ageing of UV-induced photochemistry. The~Physique des Interactions Ioniques et Moléculaires (PIIM) laboratory at Aix-Marseille Université has been conducting studies of the photochemical evolution of Titan-relevant organic ices for over a decade. The~foundational work for these measurements was established through the VUV-induced photosynthesis of organic molecules trapped in solid argon matrices at low temperatures \citep{Coupeaud2006,Couturier-Tamburelli2014SynthesisArgon}. In~\citep{Couturier-Tamburelli2014SynthesisArgon}, the~authors used a microwave-driven hydrogen lamp which generated FUV photons with energies 3--10 eV and generating a photon fluence of $\sim$10\textsuperscript{15} photons cm$^2$ s$^{-1}$ to irradiate the Ar matrix doped with a \ce{C2H2}:\ce{HC5N} mixture. The~induced photochemistry resulted in the production of \ce{HC7N} and further confirmed the possibility to generate long condensed state carbon-nitrogen chains trapped in solid Ar. Since, experiments to simulate the photochemical processing of Titan-relevant molecules at cryogenic temperatures has included \ce{C2H2} \citep{Fleury2019}, \ce{C6H6} \citep{Mouzay2020,Mouzay2021,Mouzay2021UVChemistry,Couturier-Tamburelli2025PhotochemistryAtmosphereb}, \ce{C4N2} \citep{Couturier-Tamburelli2014}, \ce{HCN}, \ce{HC3N}, and~\ce{HC5N} \citep{Couturier-Tamburelli2015,Couturier-Tamburelli2018}, and~ethyl cyanide \ce{CH3CH2CN} pure ice \citep{Couturier-Tamburelli2018BehaviourTitan}. Note in the latter case, higher energy photons (\mbox{>120 nm}) were used and yielded several organics such as ethyl isocyanide \ce{CH3CH2NC}, vinyl cyanide \ce{CH2CHCN}, and~hydrogen cyanide \ce{HCN}. Such studies not only highlight the potential for low-energy photons to initiate solid-state chemistry in lower atmospheric regions where high-energy photons are rare, but~also because they may help in retrieving desorption energies, photodissociation cross-sections, or~the branching ratios of products resulting from UV photolysis \citep{Couturier-Tamburelli2018BehaviourTitan}. In~\citep{Gudipati2013}, FUV photons at >300 nm were found to initiate condensed-state photochemistry in \ce{C4N2} ices through singlet-triplet excitations (population of triplet states only) whereas at shorter wavelengths (266 nm) kinetics are faster due to singlet-singlet excitation processes \citep{Gudipati2013,Couturier-Tamburelli2014}. Photon fluxes in these laboratory experiments reach $\sim$10\textsuperscript{17} photons cm$^2$ s$^{-1}$ which corresponds approximately to <10 Earth years on Titan depending on the experimental irradiation integration time \citep{Gudipati2013}. In~addition, with~much higher fluxes, synchrotron light is an extremely useful tool to explore a much wider range of energies; these studies are summarized in Table~\ref{Table 5}.

Several of the molecules mentioned hitherto posses their first excited singlet ($S_0-S_1$) or triplet ($S_0-T_1$) state thresholds in the FUV-VIS region which makes them particularly relevant study under cryogenic conditions, and~relevant to the cold planetary atmospheres of the outer solar system \citep{Couturier-Tamburelli2015}. As~a result, solid-state photochemistry is permitted to occur at longer wavelengths than in the gas phase by reducing the excitation energy threshold of new electronic states. In~these conditions, photochemistry initiated by low-energy (<20 eV) photons at altitudes where organic ice clouds form is plausible and may lead to pathways where ice CCN may harbor more advanced prebiotic chemistry \citep{Gudipati2013}. These processes have been extensively studied in environments applicable to the ISM (\citep{Wu2024RoleMolecules}, \textit{and references therein
}). Other ice mixtures containing more species of the CHNOPS family may for future laboratory work be relevant to incorporate and to investigate the photon- and electron-induced chemistry. For~example, the~radiolysis of ammonia (\ce{NH3}) under low-energy (7 eV) electron impact resulted in the production of the nitrogen-rich hydrazine (\ce{N2H4}) and diazene (\ce{N2H2}) \citep{Shulenberger2019Electron-InducedIces}. In~their work, incident electrons with energies as low as 6 eV permitted excition processes leading to the production of \ce{NH2} radicals at 20 K while also producing the hydride anion \ce{H-}. In~parallel to photodissociation mechanisms, electronic excitation was shown to be a crucial process in the electro-processing of \ce{NH3} ice. While ammonia remains to be directly detected on Uranus (and importantly, as~it condenses to form ammonium hydrosulfide \ce{NH4SH} clouds, see Figure~\ref{Figure 2 atm profile Uranus}), its stratospheric and tropospheric photochemical evolution (along with the other \ce{CH4} and \ce{H2S} clouds) persists as a~mystery.

\begin{table}[H]
\caption{Selected 
 branching ratios (\textit{br}) of photodissociation at Ly-$\alpha$ (121.6 nm) wavelengths (or in ranges including Ly-$\alpha$ as indicated if necessary) for \ce{CH4}, \ce{H2}, \ce{C2H2}, \ce{C2H4}, \ce{C2H6}, \ce{CH3C2H}, \ce{C3H8}, \ce{C4H2}, \ce{CH3CN}, and~\ce{H2S}. AROM indicates a summed list of 14 neutral~aromatics. \label{tab:branching-ratios-neutrals}}
\begin{tabularx}{\textwidth}{LlC}
\toprule
\textbf{Molecule} & \textbf{Photochemical Products} & \textbf{\textit{br
}} \\
\midrule
\multirow{4}{*}{CH$_4$}
  & CH$_3$ + H & 0.42 \\
  & \ce{^1CH2} + H$_2$ & 0.48 \\
  & \ce{^3CH2} + 2H & <0.1 \\
  & CH + H$_2$ + H or C + 2H$_2$ & <0.1 \\
\midrule
\multirow{2}{*}{H$_2$}
  & H$_2^*$ → H$_2$ + h$\nu'$ (fluorescence) & 0.8--0.9 \\
  & H + H (predissociation) & 0.1--0.2 \\
  \midrule
\multirow{4}{*}{\ce{C2H2}}
  & \ce{C2H + H} & 0.3 \\
  & \ce{C2 + H2} & 0.1 \\
  & \ce{C2H2^* -> C2H2} & 0.6 \\
  & \ce{C2H2+ +e-} & 0.84 $^a$ \\
  \midrule
  \multirow{2}{*}{\ce{C2H4}}
  & \ce{C2H2 + H2} & 0.58 $^b$ \\
  & \ce{C2H2 + 2H} & 0.42 $^b$ \\
\midrule
\multirow{5}{*}{\ce{C2H6}}
  & \ce{C2H4 + H2} & 0.12 \\
  & \ce{C2H4 + 2H} & 0.30 \\
  & \ce{C2H2 -> 2H2} & 0.25 \\
  & \ce{CH4 +^1CH2} & 0.25 \\
  & \ce{2CH3} & 0.08 \\
  \midrule
\multirow{2}{*}{\ce{CH3C2H}}
  & \ce{C3H3 + H} & 0.56 $^c$ \\
  & \ce{C3H2 + H2} & 0.44 $^c$ \\
   \midrule
   \multirow{4}{*}{\ce{C3H8}}
  & \ce{C3H6 + H2} & 0.34 $^d$ \\
  & \ce{C2H6 + ^1CH2} & 0.09 $^d$ \\
  & \ce{C2H5 + CH3} & 0.35 $^d$ \\
  & \ce{C2H4 + CH4} & 0.22 $^d$ \\
   \midrule
\multirow{4}{*}{\ce{C4H2}}
  & \ce{C4H + H} & 0.20 $^e$ \\
  & \ce{2C2H} & 0.03 $^e$ \\
  & \ce{C2H2 + C2} & 0.10 $^e$ \\
  & \ce{C4H2^*} & 0.67 $^e$ \\
  \midrule
  \multirow{2}{*}{\ce{CH3CN}}
  & \ce{CH3 + CN} & 0.20 $^f$ \\
  & \ce{CH2CN + H} & 0.80 $^f$ \\
  \midrule
  \multirow{1}{*}{\ce{H2S}}
  & \ce{H2 + S(^1D)} & <0.12 $^g$ \\
   \midrule
  \multirow{1}{*}{\ce{AROM}}
  & \ce{C6H6 + photoproducts} & 0.1--0.3 $^h$ \\
\bottomrule
\end{tabularx}
\noindent{\footnotesize{$^a$ From 166 to 190 nm. $^b$ From 118 to 175 nm. $^c$ Extends up to 220 nm. $^d$ From 115 to 135 nm. $^e$ From 120 to 164 nm. \mbox{$^f$ Below} 235 nm. $^g$ For $\lambda=139.11$~nm. $^h$ Model estimates from \citep{Loison2019}. Note: Although propane \ce{C3H8} has not been detected on Uranus, it has been proposed in some photochemical models. References: \citep{Wilson2004,Carrasco2007a,Carrasco2007b,Carrasco2008,Gans2013,Heays2017,Loison2019,Zhou2020UltravioletMedium,Arumainayagam2021ExtraterrestrialApplications,Chang2023ExploringMolecules,Hager2025VUVSpectra}.}}
\end{table}
\unskip

\subsection{Branching~Ratios}
\label{section branching ratios}

UV wavelength-dependent branching ratios are one of the fundamental components required to accurately simulate photochemical pathways relevant to planetary \mbox{atmospheres \citep{Gans2013}}. With~the modern advances in quantum-chemistry calculations, these important tools can provide accurate quantum yields and branching ratios for the photodissociation of neutral molecules (see Table~\ref{tab:branching-ratios-neutrals}), positive ions (Table \ref{tab:branching-ratios-ions}), and~even of vibrationally excited states of hydrocarbons of relevance \citep{Mebel2001BranchingNm}. Although~non-exhaustive, the~tables below provide wavelength-dependent branching ratios for some simple hydrocarbon photolysis reactions that have been included in Titan, Uranus (and Neptune) photochemical~models.

\begin{table}[H]
\caption{Selected quantum yields of ion-neutral and neutral-neutral reactions relevant to Titan and~Uranus. \label{tab:branching-ratios-ions}}
\begin{tabularx}{\textwidth}{LLC}
\toprule
\textbf{Reaction} & \textbf{Photochemical Pathway} & \textbf{Quantum Yield ($\Phi$)} \\
\midrule
\multirow{4}{*}{\ce{N+ + CH4}}
  & \ce{CH3+ + NH} & 0.50 \\
  & \ce{CH4+ + N} & 0.05 \\
  & \ce{H2CN+ + H2} & 0.10 \\
  & HCN+ + NH + H & 0.36 \\
  \midrule
\multirow{3}{*}{\ce{N2+ + CH4}}
  & \ce{CH2+ + N + H2} & 0.09 \\
  & \ce{CH3+ + N + H} & 0.91 \\
  & \ce{N2H+ + CH3} & - \\
  \midrule
  \multirow{1}{*}{\ce{CH3N2+}}
  & \ce{N2CH2+ + H} & 0.01 \\
    \midrule
\textbf{Atom
} & \textbf{H production channels
} & \textbf{H atom yield
} \\
    \midrule
  \multirow{5}{*}{\ce{H}}
  & \ce{CH + CH4} & 1.00 \\
  & \ce{CH + C2H6} & 0.22 \\
  & \ce{CH + C2H6} & 0.14 \\
  & \ce{CH + C3H8} & 0.19 \\
  & \ce{CH + C4H10} & 0.14 \\
\bottomrule
\end{tabularx}

\noindent{\footnotesize{References: \citep{Moses2005PhotochemistryPlanets,Loison2006RateAlkanes,Carrasco2007a,Carrasco2007b,Carrasco2008,Carrasco2008a,Carrasco2022}}}
\end{table}

\subsection{Dication Chemistry and Photo Double Ionization~Processes}

Fragmentation processes in the gas phase with respect to divalent states have recently, when investigating ionization chemistry mechanisms, gained interest. Indeed, double photoionization and photoelectron impact calculations of \ce{N2} for the first time by \citep{Lilensten2005} in Titan's upper atmosphere predicted an \ce{N2^{++}} layer located at the ionospheric peak (\mbox{$\sim$1150 km}, Reaction (\ref{double photoioniz})). Fluorescence may even be observed for \ce{N2^{++}} \citep{Lilensten2005}.
\begin{equation}
\label{double photoioniz}
\ce{A + h\nu \longrightarrow A^{++} + 2e-}
\end{equation}

The double photoionization thresholds for our atoms and molecules of interests (i.e.,~C, N, abd \ce{N2}) all occur at energies < 45 eV (34.4 eV for \ce{CH4}; the reader is referred to the detailed review by \citep{Thissen2011}, Table~\ref{Table 1 - energy range}, on~doubly charged ions in planetary atmospheres). Recently, vertical ionization pathways of bimolecular \ce{[CH4}--\ce{N2]^{2+}} or \ce{[CH4}$\cdots$\ce{CH4]^{2+}} clusters have been calculated and provided key insights into the role of monovalent and divalent states of simple hydrocarbons may play in planetary environments \citep{Thissen2011,Matsubara2014a,Matsubara2014b,Zhang2020FormationImpact}. In~\citep{Matsubara2014a}, ionized fragmentation processes led to monovalent \ce{[H3C}--\ce{HN2]+} and divalent \ce{[H4C}--\ce{N2]^{2+}} intermediates with no energy barrier. This is an important characteristic since in its ``ionized growth'', a~bimolecular cluster may thus produce a stabilized intermediate having acquired enough excess energy, making it possible to overcome energy barriers, as~described \mbox{in \citep{Matsubara2014a}}. The~calculated mechanisms also revealed the involvement of a \ce{C-N} covalent bond prior to further reactions eventually leading to the formation of the \ce{N2H+}, \ce{CH3+}, \ce{CH3N2+}, and~\ce{CH2N2+} cations. When studying \ce{[CH4}$\cdots$\ce{CH4]^{2+}} clusters, \citep{Matsubara2014b} found that the cluster would stabilize into a \ce{C-C} bond and thence produce some of the first light hydrocarbon cationic precursors. The~role divalent ionization chemistry may thus play in Titan and other planetary atmospheric chemistry is still largely unexplored but deserve further scrutiny, particularly since the double ionization thresholds of simple hydrocarbons, relevant to reduced atmospheres, are relatively small when placed in the context of upper atmospheres where higher-energy (>40 eV) electrons and photons (VUV--X-ray) precipitate. These processes underscore the need to experimentally and computationally explore ionized chemistry induced by electron and photon impact as divalent \ce{A^{++}} ionization pathways are likely to be relevant the upper planetary atmospheres as an overlooked phenomenon participating in organic growth \citep{Thissen2011}. Finally, although~at much higher electron impact energies (300 eV), the~\ce{C2H6^{2+}} dication was found to yield \ce{H3+} by vertical ionization \citep{Zhang2020FormationImpact}. Whether these processes would also occur on Uranus is at present a mystery, although~the Earth, Mars, Venus, and~Titan are all thought to harbor these dications \citep{Thissen2011}. Future constraints on these ionization processes are very relevant also in the context of data analysis, since mission mass spectrometry instruments (e.g.,~INMS and CAPS on Cassini) have not had a high enough mass resolution to distinguish between very close nuclear mass \mbox{defects \citep{Thissen2011}}. As~a result, negative ion mass spectra analysis and interpretation have, for~example, traditionally relied on assuming singly charged ions when conducting energy-to-mass data \mbox{analyses \citep{Coates2007}}. Such an assumption may be acceptable for low-mass species but loses its validity for larger molecular species and photochemical~aerosols.

\section{Negative Ion Chemistry and Haze~Growth}
\label{Section 4}
\unskip
\subsection{Anions on~Titan}


For many years, anion chemistry had remained constrained primarily to Earth's atmosphere where O (an electronegative species) atoms located in the ionosphere would undergo electron attachment reactions in the D-region, the~lower part of the \mbox{ionosphere <100 km \citep{Hargreaves1995TheEnvironment}}. Outside of Earth, negative ions have also been detected in comet Halley's inner coma (e.g., \citep{Chaizy1991NegativeHalley}), in~the Enceladus polar plume \citep{Coates2010}, and~finally on \mbox{Titan \citep{Coates2007,Coates2009,Wellbrock2013,Desai2017}}. Early photochemical models did not include anion chemistry (\mbox{e.g., \citep{Yung1984,Atreya1986}}) and it would not be until the 2007 when the discovery of very large negative ions, first up to \textit{m/z} (mass/charge) of 10,000 \citep{Coates2007} and then \textit{m/z} 13,800 by \citep{Wellbrock2013}, that our understanding of haze growth on Titan expanded and enabled photochemical models to incorporate the first steps of an unexplored chemistry \citep{Vuitton2009a,Lavvas2013}. Several anion-neutral mechanisms have been incorporated into photochemical models \citep{Vuitton2019}. While photoionization processes mainly lead to the formation of primary ions such as \ce{N2+}, \ce{CH3+}, or~H, anions can form through dissociative attachment of electrons with suprathermal \mbox{electrons \citep{Vuitton2009a,Marif2020SuprathermalIonosphere,Wang2020SuprathermalCooling} including}:
\begin{equation}
\label{}
\ce{CH4 + e- \longrightarrow CH2- + H2} ~(\text{10.3~eV})
\end{equation}
\begin{equation}
\label{}
\ce{C2H2 + e- \longrightarrow C2H- + H} ~(\text{2.7~eV})
\end{equation}

Ion-pair formation,
\begin{equation}
\label{hydride 1}
\ce{H2 + h\nu \longrightarrow H- + H+} ~(\text{17.3~eV})
\end{equation}
\begin{equation}
\label{ion pair 2}
\ce{CH4 + h\nu \longrightarrow H- + CH3+} ~(\text{21.5~eV})
\end{equation}

Radiative electron attachment with thermal electrons,
\begin{equation}
\label{}
\ce{H + e^-_{Th} \longrightarrow H- + h\nu}
\end{equation}

And proton abstraction,
\begin{equation}
\label{}
\ce{C2H2 + H- \longrightarrow C2H- + H2} ~(k=4.4 \times 10^{-9})
\end{equation}

DEA reactions are strongly dependent on accurate reaction rate coefficients and photoabsorption and photoionization cross-sections (\citep{Vuitton2019}, \textit{and references therein
}). For~most molecules considered in Titan photochemical schemes (e.g.,~\ce{HCN}, \ce{HC3N}, and~\ce{C4H2}), DEA cross-sections reach maximum values at low energies (<7 eV) \citep{Mukundan2018}. The~DEA of methane is strongly cross-section-dependent, as~seen in \citep{Mukundan2018}. Depending on the DEA cross-sections used, even one or two orders of magnitude differences can induce large variations in calculated mole fractions \citep{Mukundan2018}. Thus, reducing uncertainties in these cross-sections is fundamental since anion products will directly participate in the organics and haze growth. The~production of the simple hydride anion (Reactions (\ref{hydride 1}) and (\ref{hydride 2})), of~prime interstellar interest \citep{Millar2017}, is strongly model-dependent and an important precursor since it drives many of the proton abstraction reactions leading to the larger anionic C2, C3, and~C4 species \citep{Vuitton2009a,Martinez2010Gas-phaseH-,Mukundan2018,Nixon2024}. Thence, even with relatively low electron affinities, species such as \ce{H-} may present abundances higher than expected, considering the non-negligible \ce{CH4} abundance in the atmosphere, generally slightly less abundant than the dominant \ce{CN-}. While direct abundance measurements by Cassini of \ce{H-} were not feasible \citep{Desai2017}, the~degree of competition between DEA and proton abstraction of neutrals by \ce{H-} remains uncertain, resulting in model-observation discrepancies \citep{Dobrijevic2016}. Experimental measurements of \ce{H-} desorption from \textit{tholin
} aerosol analogues from 3 to 15 eV electron irradiation could hint at the hydride anion being a key species incorporated into the photochemical haze \citep{Pirim2015Electron-moleculeSurrogates}. Table~\ref{Table anions dominant on Titan reactions} shows the dominant anion production and loss pathways near Titan's ionospheric peak at 1100 km.
\begin{equation}
\label{hydride 2}
\ce{CH4 + e- \longrightarrow H- + CH3}
\end{equation}

Reaction pathways involving larger molecular compounds (>C3) become even more unresolved since these usually rely on slow and scant radiative attachment reactions (Reaction (\ref{RA}), see \citep{Harada2008}). The~study of these pathways has recently increased, benefiting from the detection of several anions in the ISM \citep{Millar2017}.
\begin{equation}
\label{RA}
\ce{C4H + e- \longrightarrow C4H- + h\nu}
\end{equation}

\begin{table}[H]
\caption{Dominant 
 production (first row of each species) and loss (second row of each species) mechanisms for negative ions on Titan at 1100 km inside the ionosphere. Reaction rate coefficients are given in cm$^3$ s$^{-1}$.}
\begin{tabularx}{\textwidth}{llcc}
\toprule
\textbf{Species}           & \multicolumn{1}{c}{\textbf{Reaction}}     & \multicolumn{1}{c}{\textbf{Rates} \textbf{(cm\textsuperscript{3}~s\textsuperscript{$-$1})}} & \multicolumn{1}{c}{\textbf{Ref.}} \\ \midrule

\multirow{2}{*}{\ce{H-}}                   &\ce{CH4 + e-} $\rightarrow$ \ce{H-} + \ce{CH3}  &      ($4.13 \times 10^{-11}$) &  \citep{Rawat2008AbsoluteCH4}          \\
                                      & \ce{H-} + h$\nu$ $\rightarrow$ H + e      & ($1.81 \times 10^{-2}$)      &   Miller-threshold Law         \\
                                      \midrule
\multirow{2}{*}{\ce{CN-}}                  & \ce{H-} + HCN $\rightarrow$ \ce{CN-} + \ce{H2} &  $1.50 \times 10^{-8}$     &    \citep{MackayProton-TransferExothermicity}        \\
                                      & \ce{CN-} + H $\rightarrow$ HCN + e    &  $6.30 \times 10^{-10}$     &      \citep{MackayProton-TransferExothermicity}      \\
                                      \midrule
\multirow{2}{*}{\ce{C2H-}}                 &     H$^-$ + C$_2$H$_2$ $\rightarrow$ C$_2$H$^-$ + H$_2$                            &  $3.10 \times 10^{-9}$     &    \citep{Martinez2010Gas-phaseH-}        \\
                                      &  C$_2$H$^-$ + H $\rightarrow$ \ce{C2H2} + e                              &   $1.60 \times 10^{-9}$    &   \citep{Barckholtz2001ReactionsHydrogen}         \\
                                      \midrule
\multirow{2}{*}{\ce{C3N-}}                 &  \ce{C3N} + e $\rightarrow$ \ce{C3N-} + h$\nu$                              &  $2.63 \times 10^{-10}$     &  \citep{HerbstCalculationsMedia}          \\
                                      &    \ce{C3N-} + H $\rightarrow$\ce{HC3N}                              &      $5.4 \times 10^{-10}$ &   \citep{Yang2011}         \\
                                      \midrule
\multirow{2}{*}{\ce{C5N-}}                 &  \ce{CN-} + \ce{HC5N} $\rightarrow$ \ce{C5N-} + HCN                                & $5.4 \times 10^{-9}$      &    Su-Chesnavich        \\
                                      &  \ce{C5N-} + H $\rightarrow$ \ce{HC5N}                               &      $5.8 \times 10^{-10}$ &   \citep{Yang2011}         \\
                                      \midrule
\multirow{2}{*}{\ce{C4H-}}                 &   \ce{C4H} + e $\rightarrow$ \ce{C4H-} + h$\nu$                              &  $1.1 \times 10^{-8}$     &   \citep{HerbstCalculationsMedia}         \\
                                      &   \ce{C4H-} + H $\rightarrow$ \ce{C2H2} + e                               &  $8.3 \times 10^{-9}$     &    \citep{Barckholtz2001ReactionsHydrogen}        \\
                                      \midrule
\multirow{2}{*}{\ce{C6H-}}                 &  \ce{H-} + \ce{C6H2} $\rightarrow$ \ce{C6H-} + \ce{H2}                                &  $6.3 \times 10^{-9}$     &   Langevin         \\  
                                      &    \ce{C6H-} + H $\rightarrow$ Products                           &      $5.0 \times 10^{-10}$ &    \citep{Barckholtz2001ReactionsHydrogen}        \\ 
                                      \midrule
\multirow{2}{*}{\ce{OH-}}                  &   \ce{H-} + \ce{H2O} $\rightarrow$ \ce{OH-} + \ce{H2}                               &  $4.8 \times 10^{-9}$     &     \citep{Martinez2010Gas-phaseH-}       \\
                                      &    \ce{OH-} + h$\nu$ $\rightarrow$ \ce{OH} + e                              &      $7.4 \times 10^{-3}$ & Miller-threshold Law           \\
                                      \midrule
\multirow{2}{*}{\ce{O-}}                   &   \ce{H2O} + e $\rightarrow$ \ce{O-} + \ce{H2}                               &   ($6.56 \times 10^{-12}$)    &    \citep{Itikawa2005CrossMolecules}        \\
                                      &    \ce{O-} + h$\nu$ $\rightarrow$ O + e                              &  ($1.04 \times 10^{-2}$)     & Miller-threshold Law           \\

\bottomrule       
\end{tabularx}
\label{Table anions dominant on Titan reactions}
\end{table}

Other bimolecular negative ion reactions, such as associative detachment and cation+anion reactions also exist but the latter are generally less favorable than the aforementioned pathways for the formation of anions on Titan \citep{Nixon2024}. For~a complete reaction network review, the~reader is referred to \citep{Vuitton2019}. 
Associative electron detachment have been considered in all recent photochemical models, and~rely mainly on the presence of the two most abundant radical species \citep{Vuitton2009a} in Titan's atmosphere: \ce{H} and \ce{CH3} \citep{Vuitton2009a,Vuitton2019}, following Reaction (\ref{AE}). There exists an important knowledge gap pertaining to these reactions and in particular rate constants of associative detachment reactions, and~ion-neutral reactions with H and \ce{CH3} have been assumed to be the same \citep{Vuitton2009a}. A~full bottom-up molecular growth promoted by anions is far from clear, especially given the difficulty in capturing exact ion densities with photochemical models \citep{Vuitton2019}. To~shed light on Titan's anion composition, few laboratory studies have investigated both low-mass and intermediate-mass species \citep{Horvath2009,Dubois2019ApjL}, but~it appears that nitrogenated anions (N/C > 1) could play a role in the growth of \textit{tholins
} \citep{Dubois2019ApjL,Dubois2019c}. These studies have suggested that certain molecular candidates may even contain more than three nitrogen atoms, relatively stable in N- or H-rich \mbox{environments \citep{Bierbaum2011}.}
\begin{equation}
\label{AE}
\ce{A- +B $\rightarrow$ (AB^-)^* $\rightarrow$ AB + e-}
\end{equation}

\subsection{Anions on~Uranus}

Negative ions in the atmosphere of Uranus have so far never been detected, let alone incorporated into photochemical models. In~fact, their significance has been implicitly ruled out as early as 1977 \citep{Capone1977TheNeptune}, with~the assumption of a very low abundance in the methyl radical \ce{CH3} and atomic hydrogen H (and even an exclusion of \ce{CH3} formation from the photochemical modeling of Neptune by \citep{Romani1988MethaneNeptune}), thus possibly precluding efficient electron attachment reactions on Uranus. Radical chemistry is an important mechanism for the production of negative ions. The~branching ratios and photoabsorption cross-sections of these radicals (and \ce{CH4}) are, however, either unknown or not fully characterized \citep{Pentsak2024}. As~seen on Titan previously, changes in these parameters may induce orders of magnitude variations in calculated ion densities. Furthermore, experimental work by \citep{Gans2011PhotolysisSpectrometry} have underscored the photolytic efficiency of VUV and Ly-$\alpha$ radiation to generate \ce{CH3} formation preferentially over the excited and ground-state of methylene \ce{^1CH2} and \ce{^3CH2}, respectively. These branching ratios, if~constrained accurately for the radiation conditions found at Uranus, may open a path for the formation of negative ions. Such pathways may further be possible depending on the variable solar input reaching Uranus. In~addition, DEA mechanisms may also play a central role in Uranus's atmosphere, as~was unexpectedly discovered on Titan, and~ion-pair formation triggered by photon or electron impact are also a possibility \citep{Millar2017}. In~the latter case (Reactions (\ref{hydride 1}) and (\ref{ion pair 2})), the~energy threshold is close to the parent neutral's ionization energy ($\sim$10 eV) which would make neutrals also exposed to ion-pair UV photo-destruction. Nevertheless, a~strong lack in accurate absorption cross-sections, radical recombination rates (see \citep{Bezard1999ObservationsPlanets}), and~branching ratios renders any photochemical modeling task difficult to assess whether negative ions on Uranus play an important role in the photochemical haze growth, as~is it expected on Titan. Future work in this field is much~needed. 


\subsection{Summary: Dissociative Electron Attachment (<20 eV)}

As an important source of negative ions production, DEA processes remain fundamental for photochemical models in order to calculate anion abundances \citep{Vuitton2019}. However, these depend upon accurate cross-section measurements which can sometimes vary from one experimental technique to another \citep{MaoJPhys2008}. In~addition, these cross-sections are only available for a small number of hydrocarbons and N-bearing molecules relevant to planetary atmospheres, relying at times on estimates of other similar-sized hydrocarbons \citep{MaoJPhys2008}. \mbox{Table~\ref{Table 10} lists} a condensed and non-exhaustive summary of negative ion fragments produced through low-energy DEA~processes.

\begin{table}[H]
\caption{Low-energy dissociative electron attachment of 17 parent molecules relevant to the atmospheres of Titan and~Uranus.\label{Table 10}}

\begin{adjustwidth}{-\extralength}{0cm}
\begin{tabularx}{\fulllength}{lCccC}
\toprule
\textbf{Parent} & \multicolumn{1}{c}{\textbf{Fragment}} & \multicolumn{1}{c}{\textbf{Resonance Position (eV)}} & \multicolumn{1}{c}{\textbf{Cross-Section Peak (cm\textsuperscript{2})}} & \multicolumn{1}{c}{\textbf{Ref.}} \\
\midrule
    \ce{CH4} & \ce{CH2-} & 10.4  & $1.4 \times 10^{-19}$  & \citep{Mukundan2018}     \\
     & \ce{H-} & 9.8  & $1.6 \times 10^{-18}$  & \citep{Mukundan2018} \\
    \ce{H2} &  \ce{H-} & 4.0  & $1.6 \times 10^{-21}$  & \citep{Krishnakumar2011DissociativeD2}       \\
     &   & 14  & $2.1 \times 10^{-20}$  & \citep{Krishnakumar2011DissociativeD2}       \\
     \ce{D2} & \ce{D-} & 14.0  & $5.5 \times 10^{-21}$  & \citep{Krishnakumar2011DissociativeD2}     \\
    \ce{C2H2} & \ce{C2H-} & 2.8  & $3.5 \times 10^{-20}$  & \citep{Song2017CrossAcetylene}      \\
     & \ce{C2-}  & 8.3  & $8.0 \times 10^{-21}$  & \citep{Song2017CrossAcetylene}     \\
     & \ce{H-} & 7.9  & $3.9 \times 10^{-20}$  & \citep{Song2017CrossAcetylene}      \\
     \ce{C2H4} & \ce{H-} & 10.5  & $1.9 \times 10^{-24}$  & \citep{Cadez2012LowHydrocarbons,Szymanska2014DissociativeEthylene}      \\
      & \ce{CH-} & 9.8 & \textit{ion yield
}  & \citep{Szymanska2014DissociativeEthylene}      \\
      & \ce{C2H-} & 9.8 & \textit{ion yield
}  & \citep{Szymanska2014DissociativeEthylene}      \\
      & \ce{C2H2-} & 1.6 & \textit{ion yield
}  & \citep{Szymanska2014DissociativeEthylene}      \\
      & \ce{C2H3-} & 7.0 & \textit{ion yield
}  & \citep{Szymanska2014DissociativeEthylene}      \\
    \ce{C2H6} & \ce{H-} & 9.2 & \textit{ion yield
}  & \citep{Cadez2012LowHydrocarbons}      \\
    \ce{C3H4} & \ce{C3H3-} & 3.4  & $1.9 \times 10^{-24}$  & \citep{Janeckova2012DissociativeProximity}      \\
    \ce{C3H8} & \ce{H-} & 8.6 & \textit{ion yield
}  & \citep{Cadez2012LowHydrocarbons}      \\
    \ce{C4H2} & \ce{C4H-} & 2.5  & $3.0 \times 10^{-24}$  & \citep{May2008AbsoluteDiacetylene}      \\
     &  & 5.3  & $7.3 \times 10^{-23}$  & \citep{May2008AbsoluteDiacetylene}      \\
    \ce{C4H6} & \ce{H-} & 4.0  & $1.4 \times 10^{-24}$  & \citep{Janeckova2012DissociativeProximity}      \\
    \ce{C6H2} & \ce{C6H-} & 2.8  & $3.5 \times 10^{-20}$,~\textit{est.
}  & \citep{MaoJPhys2008,May2008AbsoluteDiacetylene,Vuitton2009a}      \\
    \ce{HCN} &  \ce{CN-} & 1.9  & $9.4 \times 10^{-22}$  & \citep{May2010AbsoluteDCN}      \\
    \ce{DCN} &  \ce{CN-} & 1.9  & $3.4 \times 10^{-22}$  & \citep{May2010AbsoluteDCN}      \\
    \ce{NH3} &  \ce{H-} & 5.7  & $2.3 \times 10^{-18}$  &  \citep{Rawat2008AbsoluteCH4}     \\
     &  \ce{NH2-} & 5.9  & $1.6 \times 10^{-18}$  &  \citep{Rawat2008AbsoluteCH4}     \\
    \ce{CH2N2} &  \ce{CN-} & 6.4  & $3.9 \times 10^{-20}$  &  \citep{Tanzer2015}     \\
    \ce{C2H4N2} &  \ce{CN-} & 1.9  & $3.9 \times 10^{-20}$  &  \citep{Pelc2016FormationAminoacetonitrile}     \\
    \ce{H2S} &  \ce{HS-} & 1.6  & $1.8 \times 10^{-18}$  &  \citep{Fiquet-Fayard1972FormationD2S,Rao1993ElectronH2S}     \\
     &  \ce{S-} & 9.7  & $4.4 \times 10^{-19}$  &  \citep{Fiquet-Fayard1972FormationD2S,Rao1993ElectronH2S}     \\
    \bottomrule
\end{tabularx}
\end{adjustwidth}
\noindent{\footnotesize{Note: Certain experiments have provided the ion yields of fragments and these are annotated here. More DEA cross-section data can be found at the Innsbruck Dissociative Electron Attachment DataBase (IDEADB) node: \url{https://ideadb.uibk.ac.at/}, accessed on 9 June 2025
.}}
\end{table}

\section{Summary: Opportunities for Future~Studies}
\label{Section 5 conclusions}

The planetary atmospheres of the cold regions of the outer solar system provide a unique natural laboratory presenting a wide array of chemical compositions and atmosphere-magnetosphere interactions. The~composition of these atmospheres is also prone to the variables of seasonal changes and exogenic material precipitation, both affecting their thermal, chemical, and~physical characteristics. The~succession of the Pioneer, Voyager, and~Cassini-Huygens missions opened new pathways towards the characterization of the gas giants while simultaneously spawning even more questions on their origins and evolution. Uranus and Neptune, the~two least explored planets in the solar system, likely portray features that will ultimately improve our understanding of the formation of our solar system and that of exoplanets, too. The~Cassini mission in particular has unearthed an invaluable amount of discovery at Titan and will continue to provide the community with an ameliorated description of the global atmosphere, given of wealth of non-analyzed data. With~Uranus being placed at a top priority for a future flagship \mbox{mission \citep{Decadal20232032}}, the~available breadth of observational, theoretical, and~experimental techniques will help support the future investigation of the Uranian environment. As~detailed in a recent community input poll, the~interest in future Uranian studies encompasses a large scope of domains, from~the study of the atmosphere, rings and satellites, the~interior, and~the magnetosphere \citep{Simon2025UranusPoll}. This review attempts to describe the broad effects of photochemical processes induced by low-energy (<50 eV) photons and electrons on the growth of organic molecules in the gas and solid phases. Future synergistic work can help address many of the scientific questions and uncertainties laid out here, in~order to better understand the formation mechanisms of organic molecules and chemically complex haze~particles.

\vspace{6pt}

\funding{NASA SMD support for this work is~acknowledged.}

\acknowledgments{The author expresses gratitude to two anonymous reviewers who contributed to the improvement of this review through constructive remarks, and~to the Guest Editor of this Special Issue, Hassan~Abdoul-Carime. 
 The author gratefully acknowledges insightful discussions with Lora Jovanovic and her assistance with figure improvements. Sincere appreciation is also extended to Benoît Seignovert, Nathalie Carrasco, Olivier Mousis, Mark Hofstadter, Michel Nuevo, Jean-Pierre Lebreton
, Toshiaki Matsubara, Salma Bejaoui
, Ella Sciamma-O'Brien, and~Farid Salama for valuable scientific exchanges. The~author is particularly thankful to Chris Arumainayagam for helpful guidance regarding low-energy electron interactions, and~to Partha Bera for guidance in understanding vertical ionization processes. Special thanks are due to Christiaan Boersma for his consistent encouragement. The~author also wishes to thank Mustapha Meftah for generously providing the high-resolution SOLAR-HRS SSI raw data. Deep gratitude is expressed to Hadir Marei for her foundational and ongoing motivation throughout the preparation of this work. Any oversights or errors remain the sole responsibility of the author. This work has made use of NASA’s Astrophysics Data System. S.D.G.}


\conflictsofinterest{The author declares no conflicts of~interest.} 



\abbreviations{Abbreviations}{

\vspace{-3pt}
\noindent 
\begin{tabular}{@{}ll}
UOP & Uranus Orbiter Probe\\
GCR & Galactic Cosmic Rays\\
UVS & Ultraviolet Spectrometer\\
NUV & Near Ultraviolet\\
FUV & Far Ultraviolet\\
VUV & Vacuum Ultraviolet\\
EUV & Extreme Ultraviolet\\
LIPM & Local Interplanetary Medium\\
ISRF & Interstellar Radiation Field\\
ISM & Interstellar Medium\\
JWST & James Webb Space Telescope\\
CRIR & Cosmic Ray Ionization Rate\\
INMS & Ion and Neutral Mass Spectrometer\\
CAPS & Cassini Plasma Spectrometer\\
SSI & Solar Spectrum Irradiance\\
CCN & Cloud Condensation Nuclei\\
DEA & Dissociative Electron Attachment\\
TNI & Temporary Negative Ions
\end{tabular}
}




\begin{adjustwidth}{-\extralength}{0cm}

\reftitle{References}

\PublishersNote{}
\end{adjustwidth}

\end{document}